\documentclass[sigconf]{acmart}
\AtBeginDocument{%
  }

\copyrightyear{2026}
\acmYear{2026}
\setcopyright{cc}
\setcctype{by}
\acmConference[CCS '26] {Proceedings of the 2026 ACM SIGSAC Conference on Computer and Communications Security}{November 15--19, 2026}{The Hague, Netherlands.}
\acmBooktitle{Proceedings of the 2026 ACM SIGSAC Conference on Computer and Communications Security (CCS '26), November 15--19, 2026, The Hague, Netherlands}
\acmISBN{979-8-4007-2871-6/2026/11}
\acmDOI{10.1145/3830454.3846554}

\makeatletter
\g@addto@macro\@copyrightpermission{\par This is a corrected version (v2) of the paper originally published in:}
\makeatother

\usepackage[nohyperlinks, printonlyused, withpage, nolist]{acronym}
\usepackage{numprint}
\usepackage{threeparttable}
\npthousandthpartsep{,}
\npdecimalsign{.}
\usepackage{xspace}
\usepackage[capitalize,sort&compress]{cleveref}
\crefname{figure}{Figure}{Figures}
\Crefname{figure}{Figure}{Figures}
\usepackage{pgf}
\usepackage{tikz}
\usetikzlibrary{positioning,arrows.meta}
\usepackage{import}
\usepackage{multirow}
\usepackage{siunitx}

\everymath=\expandafter{\the\everymath\displaystyle}

\newcommand{\PctPrint}[1]{\numprint{#1}\,\%\xspace}

\newcommand{\TrainTrackJourneys}{\numprint{217}\xspace}
\newcommand{\TrainTrackTrips}{\numprint{342}\xspace}
\newcommand{\TrainTrackSegments}{\numprint{1979}\xspace}

\newcommand{\TrainTrackStations}{\numprint{246}\xspace}
\newcommand{\TrainTrackRecordingDuration}{\numprint{482.6}\xspace}
\newcommand{\TrainTrackLongDistanceDuration}{\numprint{414.7}\xspace}
\newcommand{\TrainTrackLongDistanceSegments}{\numprint{844}\xspace}
\newcommand{\TrainTrackRegionalDuration}{\numprint{44.1}\xspace}
\newcommand{\TrainTrackRegionalSegments}{\numprint{342}\xspace}
\newcommand{\TrainTrackUrbanDuration}{\numprint{19.7}\xspace}
\newcommand{\TrainTrackUrbanSegments}{\numprint{586}\xspace}
\newcommand{\TrainTrackMetroDuration}{\numprint{4.2}\xspace}
\newcommand{\TrainTrackMetroSegments}{\numprint{201}\xspace}
\newcommand{\TrainTrackTramDuration}{\numprint{0.1}\xspace}
\newcommand{\TrainTrackTramSegments}{\numprint{6}\xspace}

\newcommand{\TrainTrackElectricYesDuration}{\numprint{443.2}\xspace}
\newcommand{\TrainTrackElectricYesCount}{\numprint{1189}\xspace}

\newcommand{\TrainTrackLongDistanceElectricYesDuration}{\numprint{402.6}\xspace}
\newcommand{\TrainTrackLongDistanceElectricYesCount}{\numprint{822}\xspace}

\newcommand{\TrainTrackRegionalElectricYesDuration}{\numprint{37.0}\xspace}
\newcommand{\TrainTrackRegionalElectricYesCount}{\numprint{279}\xspace}

\newcommand{\TrainTrackUrbanElectricYesDuration}{\numprint{3.7}\xspace}
\newcommand{\TrainTrackUrbanElectricYesCount}{\numprint{88}\xspace}

\newcommand{\TrainTrackMetroElectricYesDuration}{\numprint{0.0}\xspace}
\newcommand{\TrainTrackMetroElectricYesCount}{\numprint{0}\xspace}

\newcommand{\TrainTrackTramElectricYesDuration}{\numprint{0.0}\xspace}
\newcommand{\TrainTrackTramElectricYesCount}{\numprint{0}\xspace}

\newcommand{\TrainTrackLongDistanceElectricYesCountPct}{\PctPrint{97.4}}

\newcommand{\TrainTrackLongDistanceElectricYesDurationPct}{\PctPrint{97.1}}

\newcommand{\TrainTrackRegionalElectricYesCountPct}{\PctPrint{81.6}}

\newcommand{\TrainTrackRegionalElectricYesDurationPct}{\PctPrint{83.9}}

\newcommand{\TrainTrackUrbanElectricYesCountPct}{\PctPrint{15.0}}

\newcommand{\TrainTrackUrbanElectricYesDurationPct}{\PctPrint{18.9}}

\newcommand{\TrainTrackMetroElectricYesCountPct}{\PctPrint{0.0}}

\newcommand{\TrainTrackMetroElectricYesDurationPct}{\PctPrint{0.0}}

\newcommand{\TrainTrackTramElectricYesCountPct}{\PctPrint{0.0}}

\newcommand{\TrainTrackTramElectricYesDurationPct}{\PctPrint{0.0}}

\newcommand{\TrainTrackElectricYesCountPct}{\PctPrint{60.1}}

\newcommand{\TrainTrackElectricYesDurationPct}{\PctPrint{91.8}}

\newcommand{\TmdTrainActivityTime}{\numprint{1737481}\,s\xspace}

\newcommand{\ColocDatasetAllSegmentsMax}{\numprint{357}\xspace}

\newcommand{\ColocDatasetAllSegmentsFirstTen}{\numprint{348}\xspace}

\newcommand{\ColocDatasetLongDistanceSegmentsFirstThreeHundred}{\numprint{197}\xspace}

\newcommand{\ColocDatasetRegionalSegmentsFirstThreeHundred}{\numprint{39}\xspace}
\newcommand{\ColocDatasetRegionalSegmentsFirstSixHundred}{\numprint{10}\xspace}
\newcommand{\ColocDatasetRegionalSegmentsFirstNineHundred}{\numprint{2}\xspace}

\newcommand{\ColocDistanceDTWBestMCC}{\numprint{0.60}\xspace}

\newcommand{\ColocDistanceAbsDDTWBestFFirstSixty}{\numprint{0.02}$\pm$\numprint{0.01}\xspace}

\newcommand{\ColocDistanceAbsDDTWBestFFirstThreeHundred}{\numprint{0.02}$\pm$\numprint{0.03}\xspace}

\newcommand{\ColocDistanceAbsDDTWBestFFirstSixHundred}{\numprint{0.02}$\pm$\numprint{0.01}\xspace}

\newcommand{\ColocDistanceAbsDDTWBestFFirstNineHundred}{\numprint{0.02}$\pm$\numprint{0.01}\xspace}

\newcommand{\ColocDistanceAbsDDTWBestMCCFirstSixty}{\numprint{0.09}$\pm$\numprint{0.05}\xspace}

\newcommand{\ColocDistanceAbsDDTWBestMCCFirstThreeHundred}{\numprint{0.06}$\pm$\numprint{0.08}\xspace}

\newcommand{\ColocDistanceAbsDDTWBestMCCFirstSixHundred}{\numprint{0.01}$\pm$\numprint{0.02}\xspace}

\newcommand{\ColocDistanceAbsDDTWBestMCCFirstNineHundred}{\numprint{0.02}$\pm$\numprint{0.03}\xspace}

\newcommand{\ColocDistanceDTWROneBestFFirstSixty}{\numprint{0.13}$\pm$\numprint{0.05}\xspace}

\newcommand{\ColocDistanceDTWROneBestFFirstThreeHundred}{\numprint{0.38}$\pm$\numprint{0.05}\xspace}

\newcommand{\ColocDistanceDTWROneBestFFirstSixHundred}{\numprint{0.42}$\pm$\numprint{0.11}\xspace}

\newcommand{\ColocDistanceDTWROneBestFFirstNineHundred}{\numprint{0.56}$\pm$\numprint{0.12}\xspace}

\newcommand{\ColocDistanceDTWROneBestMCCFirstSixty}{\numprint{0.17}$\pm$\numprint{0.04}\xspace}

\newcommand{\ColocDistanceDTWROneBestMCCFirstThreeHundred}{\numprint{0.43}$\pm$\numprint{0.05}\xspace}

\newcommand{\ColocDistanceDTWROneBestMCCFirstSixHundred}{\numprint{0.49}$\pm$\numprint{0.09}\xspace}

\newcommand{\ColocDistanceDTWROneBestMCCFirstNineHundred}{\numprint{0.60}$\pm$\numprint{0.11}\xspace}

\newcommand{\ColocDistanceDTWROneMCCFirstSixtySRTen}{\numprint{0.13}$\pm$\numprint{0.04}\xspace}

\newcommand{\ColocDistanceDTWROneMCCFirstSixtySRTwenty}{\numprint{0.15}$\pm$\numprint{0.07}\xspace}

\newcommand{\ColocDistanceDTWROneMCCFirstSixtySRFourty}{\numprint{0.14}$\pm$\numprint{0.02}\xspace}

\newcommand{\ColocDistanceDTWROneMCCFirstSixtySRSixty}{\numprint{0.17}$\pm$\numprint{0.04}\xspace}

\newcommand{\ColocDistanceDTWROneMCCFirstThreeHundredSRTen}{\numprint{0.31}$\pm$\numprint{0.03}\xspace}

\newcommand{\ColocDistanceDTWROneMCCFirstThreeHundredSRTwenty}{\numprint{0.31}$\pm$\numprint{0.03}\xspace}

\newcommand{\ColocDistanceDTWROneMCCFirstThreeHundredSRFourty}{\numprint{0.33}$\pm$\numprint{0.04}\xspace}

\newcommand{\ColocDistanceDTWROneMCCFirstThreeHundredSRSixty}{\numprint{0.43}$\pm$\numprint{0.05}\xspace}

\newcommand{\ColocDistanceDTWROneMCCFirstSixHundredSRTen}{\numprint{0.26}$\pm$\numprint{0.05}\xspace}

\newcommand{\ColocDistanceDTWROneMCCFirstSixHundredSRTwenty}{\numprint{0.25}$\pm$\numprint{0.06}\xspace}

\newcommand{\ColocDistanceDTWROneMCCFirstSixHundredSRFourty}{\numprint{0.26}$\pm$\numprint{0.06}\xspace}

\newcommand{\ColocDistanceDTWROneMCCFirstSixHundredSRSixty}{\numprint{0.49}$\pm$\numprint{0.09}\xspace}

\newcommand{\ColocDistanceDTWROneMCCFirstNineHundredSRTen}{\numprint{0.30}$\pm$\numprint{0.04}\xspace}

\newcommand{\ColocDistanceDTWROneMCCFirstNineHundredSRTwenty}{\numprint{0.29}$\pm$\numprint{0.04}\xspace}

\newcommand{\ColocDistanceDTWROneMCCFirstNineHundredSRFourty}{\numprint{0.27}$\pm$\numprint{0.05}\xspace}

\newcommand{\ColocDistanceDTWROneMCCFirstNineHundredSRSixty}{\numprint{0.60}$\pm$\numprint{0.11}\xspace}

\newcommand{\ColocDistanceLongDistanceDTWROneBestMCCFirstSixty}{\numprint{0.18}$\pm$\numprint{0.06}\xspace}

\newcommand{\ColocDistanceLongDistanceDTWROneBestMCCFirstThreeHundred}{\numprint{0.40}$\pm$\numprint{0.11}\xspace}

\newcommand{\ColocDistanceLongDistanceDTWROneBestMCCFirstSixHundred}{\numprint{0.44}$\pm$\numprint{0.24}\xspace}

\newcommand{\ColocDistanceLongDistanceDTWROneBestMCCFirstNineHundred}{\numprint{0.58}$\pm$\numprint{0.14}\xspace}

\newcommand{\ColocDistanceRegionalDTWROneBestMCCFirstSixty}{\numprint{0.43}$\pm$\numprint{0.06}\xspace}

\newcommand{\ColocDistanceRegionalDTWROneBestMCCFirstThreeHundred}{\numprint{0.58}$\pm$\numprint{0.08}\xspace}

\newcommand{\MlResultBestFOverall}{\numprint{0.90}\xspace}
\newcommand{\MlResultBestFDuration}{900\xspace}

\newcommand{\MlResultBestFWindow}{5\xspace}

\newcommand{\MlResultBestFChunkSixtyDurationSixty}{\numprint{0.84}\xspace}
\newcommand{\MlResultBestFChunkSixtyDurationThreeHundred}{\numprint{0.85}\xspace}
\newcommand{\MlResultBestFChunkSixtyDurationSixHundred}{\numprint{0.86}\xspace}
\newcommand{\MlResultBestFChunkSixtyDurationNineHundred}{\numprint{0.90}\xspace}
\newcommand{\MlResultBestMCCOverall}{\numprint{0.79}\xspace}

\newcommand{\MlResultBestMCCChunk}{60\xspace}

\newcommand{\MlResultBestMCCHz}{60\xspace}
\newcommand{\MlResultBestMCCChunkSixtyDurationSixty}{\numprint{0.69}\xspace}
\newcommand{\MlResultBestMCCChunkSixtyDurationThreeHundred}{\numprint{0.73}\xspace}
\newcommand{\MlResultBestMCCChunkSixtyDurationSixHundred}{\numprint{0.73}\xspace}
\newcommand{\MlResultBestMCCChunkSixtyDurationNineHundred}{\numprint{0.79}\xspace}
\newcommand{\MlResultBestMCCFirstSixty}{\numprint{0.69}\xspace}
\newcommand{\MlResultBestMCCFirstThreeHundred}{\numprint{0.73}\xspace}
\newcommand{\MlResultBestMCCFirstSixHundred}{\numprint{0.73}\xspace}
\newcommand{\MlResultBestMCCFirstNineHundred}{\numprint{0.79}\xspace}
\newcommand{\MlResultLongDistanceBestMCCFirstSixty}{\numprint{0.65}\xspace}
\newcommand{\MlResultLongDistanceBestMCCFirstThreeHundred}{\numprint{0.70}\xspace}
\newcommand{\MlResultLongDistanceBestMCCFirstSixHundred}{\numprint{0.72}\xspace}
\newcommand{\MlResultLongDistanceBestMCCFirstNineHundred}{\numprint{0.80}\xspace}

\newcommand{\MlResultRegionalBestMCCFirstSixty}{\numprint{0.92}\xspace}
\newcommand{\MlResultRegionalBestMCCFirstThreeHundred}{\numprint{0.85}\xspace}
\newcommand{\MlResultRegionalBestMCCFirstSixHundred}{\numprint{0.91}\xspace}
\newcommand{\MlResultRegionalBestMCCFirstNineHundred}{\numprint{0.74}\xspace}
\newcommand{\MlResultRegionalBestMCCOverall}{\numprint{0.92}\xspace}

\newcommand{\TmdResultBestMccFirstAllDTenWOne}{\numprint{0.64}\xspace}

\newcommand{\TmdResultBestMccFirstAllDTenWThree}{\numprint{0.72}\xspace}

\newcommand{\TmdResultBestMccFirstAllDTenWSix}{\numprint{0.81}\xspace}

\newcommand{\TmdResultBestMccFirstAllDTenWNine}{\numprint{0.86}\xspace}

\newcommand{\TmdResultBestMccFirstAllDTenWTwelve}{\numprint{0.88}\xspace}

\newcommand{\TmdResultBestMccFirstSixtyDTenWOne}{\numprint{0.78}\xspace}

\newcommand{\TmdResultBestMccFirstSixtyDTenWThree}{\numprint{0.83}\xspace}

\newcommand{\TmdResultBestMccFirstSixtyDTenWSix}{\numprint{0.85}\xspace}

\newcommand{\TmdResultBestMccFirstSixtyDTenWNine}{\numprint{0.86}\xspace}

\newcommand{\TmdResultBestMccFirstSixtyDTenWTwelve}{\numprint{0.87}\xspace}

\newcommand{\TmdResultBestMccFirstThreeHundredDTenWOne}{\numprint{0.83}\xspace}

\newcommand{\TmdResultBestMccFirstThreeHundredDTenWThree}{\numprint{0.87}\xspace}

\newcommand{\TmdResultBestMccFirstThreeHundredDTenWSix}{\numprint{0.91}\xspace}

\newcommand{\TmdResultBestMccFirstThreeHundredDTenWNine}{\numprint{0.92}\xspace}

\newcommand{\TmdResultBestMccFirstThreeHundredDTenWTwelve}{\numprint{0.93}\xspace}

\newcommand{\TmdResultBestMccFirstSixHundredDTenWOne}{\numprint{0.80}\xspace}

\newcommand{\TmdResultBestMccFirstSixHundredDTenWThree}{\numprint{0.85}\xspace}

\newcommand{\TmdResultBestMccFirstSixHundredDTenWSix}{\numprint{0.91}\xspace}

\newcommand{\TmdResultBestMccFirstSixHundredDTenWNine}{\numprint{0.92}\xspace}

\newcommand{\TmdResultBestMccFirstSixHundredDTenWTwelve}{\numprint{0.93}\xspace}

\newcommand{\TmdResultBestMccFirstNineHundredDTenWOne}{\numprint{0.78}\xspace}

\newcommand{\TmdResultBestMccFirstNineHundredDTenWThree}{\numprint{0.83}\xspace}

\newcommand{\TmdResultBestMccFirstNineHundredDTenWSix}{\numprint{0.88}\xspace}

\newcommand{\TmdResultBestMccFirstNineHundredDTenWNine}{\numprint{0.90}\xspace}

\newcommand{\TmdResultBestMccFirstNineHundredDTenWTwelve}{\numprint{0.91}\xspace}

\newcommand{\TmdResultBestMccOneChunk}{\numprint{0.83}\xspace}
\newcommand{\TmdResultBestFOneChunk}{\numprint{0.94}\xspace}
\newcommand{\TmdResultBestMccOverall}{\numprint{0.93}\xspace}
\newcommand{\TmdResultBestFOverall}{\numprint{0.98}\xspace}

\newcommand{\RuntimeDtwMin}{\SI{1.1}{\milli\second}\xspace}
\newcommand{\RuntimeDtwMax}{\SI{527.1}{\milli\second}\xspace}
\newcommand{\RuntimeMlMin}{\SI{3.2}{\milli\second}\xspace}
\newcommand{\RuntimeMlMax}{\SI{26.3}{\milli\second}\xspace}

\newcommand{\RuntimeMlPerChunkAmortised}{\SI{0.7}{\milli\second}\xspace}

\newcommand{\RuntimeSpeedupMean}{\numprint{9}\xspace}
\newcommand{\RuntimeSpeedupMax}{\numprint{22}\xspace}

\begin{document}

\title[On the Same Track: Privacy Leaks in Electric Rail Transport via Magnetic Fields]{On the Same Track: Privacy Leaks in\\ Electric Rail Transport via Magnetic Fields}

\author{Stefan Czybik}
\authornotemark[1]
\orcid{/0009-0001-3342-7460}
\affiliation{%
  \institution{BIFOLD \& TU Berlin}
  \city{Berlin}
  \country{Germany}
}

\author{Pia Hanfeld}
\authornotemark[1]
\orcid{/0009-0009-8198-5764}
\affiliation{%
  \institution{BIFOLD \& TU Berlin}
  \city{Berlin}
  \country{Germany}
}

\author{Konrad Rieck}
\orcid{/0000-0002-5054-8758}
\affiliation{%
  \institution{BIFOLD \& TU Berlin}
  \city{Berlin}
  \country{Germany}
}

\renewcommand{\shortauthors}{Stefan Czybik, Pia Hanfeld, \& Konrad Rieck }

\begin{abstract}
Inertial sensors of smartphones are a well-known privacy risk, as they allow inferring sensitive user movements and activity without explicit permissions.
Among these sensors, the magnetometer has received little attention, as it is assumed to measure mainly the Earth's magnetic field.
In this paper, we show that electric trains generate distinctive magnetic fields that can be captured by a smartphone magnetometer.
Apps deployed on the device can access these signals without user permission, enabling two novel attacks on passenger privacy:
First, we are able to determine whether the user is riding an electric train (transport mode detection).
Second, we can infer whether other victims are aboard the same train due to shared patterns in the magnetic field (colocation prediction).
In an empirical evaluation across \TrainTrackJourneys~train rides, we demonstrate that both attacks are effective, confirming a real-world threat.
We discuss countermeasures and argue that magnetometer access should require explicit user permission in mobile operating systems.
\end{abstract}
\acresetall



\maketitle

\begin{acronym}[ECU]
    \acro{TMD}[TMD]{transport mode detection}
    \acro{TRD}[TRD]{train ride detection}
    \acro{SHL}[SHL]{Sussex-Huawei Locomotion–Transportation}
    \acro{MEMS}[MEMS]{micro-electromechanical system}
    \acro{LSTM}[LSTM]{long short-term memory}
    \acro{CNN}[CNN]{convolutional neural network}
    \acro{ML}[ML]{machine learning}
    \acro{AC}[AC]{alternating current}
    \acro{DC}[DC]{direct current}
    \acro{DTW}[DTW]{dynamic time warping}
    \acro{DDTW}[DDTW]{derivative dynamic time warping}
    \acro{PSD}[PSD]{power spectral density}
    \acro{FFT}[FFT]{fast Fourier transform}
    \acro{DFT}[DFT]{discrete Fourier transform}
    \acro{SNR}[SNR]{signal-to-noise ratio}
    \acro{XGBoost}[XGBoost]{extreme gradient boosting}
    \acro{RSPD}[RSPD]{robust spectral peak detection}
    \acro{NSGAII}[NSGA-II]{nondominated sorting genetic algorithm II}
    \acro{CFAR}[CFAR]{constant false alarm rate}
    \acro{MCC}[MCC]{Matthews correlation coefficient}
\end{acronym}

\section{Introduction}

Inertial sensors are an integral component of modern smartphones. 
Composed of an accelerometer, gyroscope, and magnetometer, these sensors continuously track the orientation and motion of the device during use. 
Such measurements are essential for a wide range of everyday functions, including screen rotation, step counting, gesture recognition, navigation assistance, and image stabilization~\citep{lane2010survey}. 
Because these functions are pervasive across the mobile ecosystem, neither Android nor iOS protects inertial sensors behind explicit permissions, allowing any installed application to access their readings without notice~\citep{delgado-santosSurveyPrivacyVulnerabilities2022}.
This unrestricted access has repeatedly been shown to pose privacy risks. 

{\renewcommand\thefootnote{\ensuremath{*}}%
 \footnotetext{Authors contributed equally.}}

Prior work has demonstrated that inertial sensors can reveal various sensitive information about the user~\citep{delgado-santosSurveyPrivacyVulnerabilities2022}, ranging from traveled routes~\citep{NarVo-BloNou+16,huaWeCanTrack2017,kouam2026feasibility} and transportation mode~\citep{fengTransportationModeRecognition2013} to typed text~\citep{MarVerCarTra+11,owusu2012accessory} and even spoken words~\citep{MicBonNak14,HuZhuSanSpo+22}. 
Within this body of work, however, the magnetometer has received comparatively little attention so far. 
It is commonly assumed to measure mainly the Earth's static magnetic field, providing a compass heading of limited use to an adversary.
This is because, at first glance, other magnetic fields strong enough to be captured by a smartphone magnetometer seem rare in everyday environments.

In this paper, we challenge this assumption. We show that electric trains, powered by alternating current, generate distinctive magnetic fields that are measurable inside the passenger cabin.
As electric current flows through the overhead power line and traction motors, it produces a magnetic emission that penetrates the train's metal shell into the passenger cabin.
The emission varies with acceleration, route, and train load, forming a fingerprint of the ride.
We find that this fingerprint can be captured by common smartphone magnetometers, opening a new channel for inferring sensitive information about passengers.

Building on this observation, we introduce two attacks on privacy.
First, we present a new variant of \emph{transport mode detection}: we determine whether the user is riding an electric train from the magnetometer alone. 
This is possible because the electric current oscillates at the rail grid frequency, producing a distinctive magnetic signature.
Second, we present \emph{colocation prediction}: by comparing magnetic field traces between smartphones, we are able to infer whether two users are traveling on the same train. 
The attack rests on the fact that passengers in the same train are immersed in the same magnetic field, so their traces share a synchronous pattern.
Both attacks can be conducted covertly on Android and iOS, since magnetometer access is unrestricted on both platforms.

We empirically evaluate the attacks on a dataset of \TrainTrackJourneys train rides collected across diverse routes, covering long-distance and regional services. 
For transport mode detection, we achieve a \ac{MCC} score of \TmdResultBestMccOverall (F1: \TmdResultBestFOverall), on par with the state of the art.
Our method, however, is notably simpler than prior work, as detecting the distinct frequency of the magnetic field requires little feature engineering.
For colocation inference, we achieve an \ac{MCC} score of \MlResultBestMCCOverall (F1: \MlResultBestFOverall) using a learning-based approach, outperforming prior methods based on other sensors.
Our results demonstrate that magnetometer-based inference is a real-world privacy threat, complementing existing attack vectors. 

Both attacks apply specifically to electric trains operating on alternating current (AC) and thus not to diesel routes or DC-powered subway lines. 
However, AC traction dominates long-distance and regional rail across Europe and Asia, exposing a substantial passenger population to the threats we describe. 
More broadly, our findings add to the growing evidence that unrestricted access to inertial sensors continues to expose users to privacy threats. 
We discuss possible countermeasures for the magnetometer in this work, but ultimately join recent calls to systematically restrict access to these sensors in mobile operating systems.

\smallskip\noindent
In summary, we make the following contributions in this work:
\begin{itemize}
     \setlength{\itemsep}{3pt}
     \item \textbf{Magnetic fields of electric trains.} We identify the magnetic field of electric trains as a previously overlooked privacy-relevant signal, readily captured by common smartphone magnetometers inside the passenger cabin.
    
    \item \textbf{Novel privacy attacks.} We present two novel attacks based on magnetic fields in trains that enable transport mode detection and colocation inference of passengers from the magnetometer alone.
    
    \item \textbf{Empirical evaluation and dataset.} We demonstrate the efficacy of both attacks in real-world conditions across \TrainTrackJourneys~train rides, showing that electric train rides and jointly traveling passengers can be identified with high accuracy.
\end{itemize}

To foster research on privacy in rail transport, we publicly release our implementation along with our dataset, which to the best of our knowledge constitutes the largest collection of train rides with recorded inertial sensor data.

\section{Magnetic Fields in Trains}
\label{sec:background}

Our attacks build on three pieces of background: the physics of magnetic fields, how electric trains produce distinctive magnetic emissions through their traction systems, and how smartphone magnetometers capture these signals. 

\subsection{Magnetic Fields}

Magnetic fields arise from two fundamental sources: electric currents and the intrinsic magnetic moments of materials such as permanent magnets.
The most familiar example in everyday life is the Earth's magnetic field, generated by electric currents in the planet's molten outer core. 
Near the surface, its magnitude is on the order of tens of microtesla, and it is typically treated as a static background field that supports compass-based navigation.

Beyond natural sources, magnetic fields are generated by electric currents, since every current-carrying conductor produces a field around it.
Such fields are commonly classified as either static or alternating. Static fields are produced by \ac{DC} and remain constant over time. 
Alternating fields, in contrast, are generated by \ac{AC} and change periodically with the current's frequency. 
In practice, magnetic fields produced by alternating currents are thus associated with a characteristic frequency, which simplifies their detection and analysis, as we discuss later.

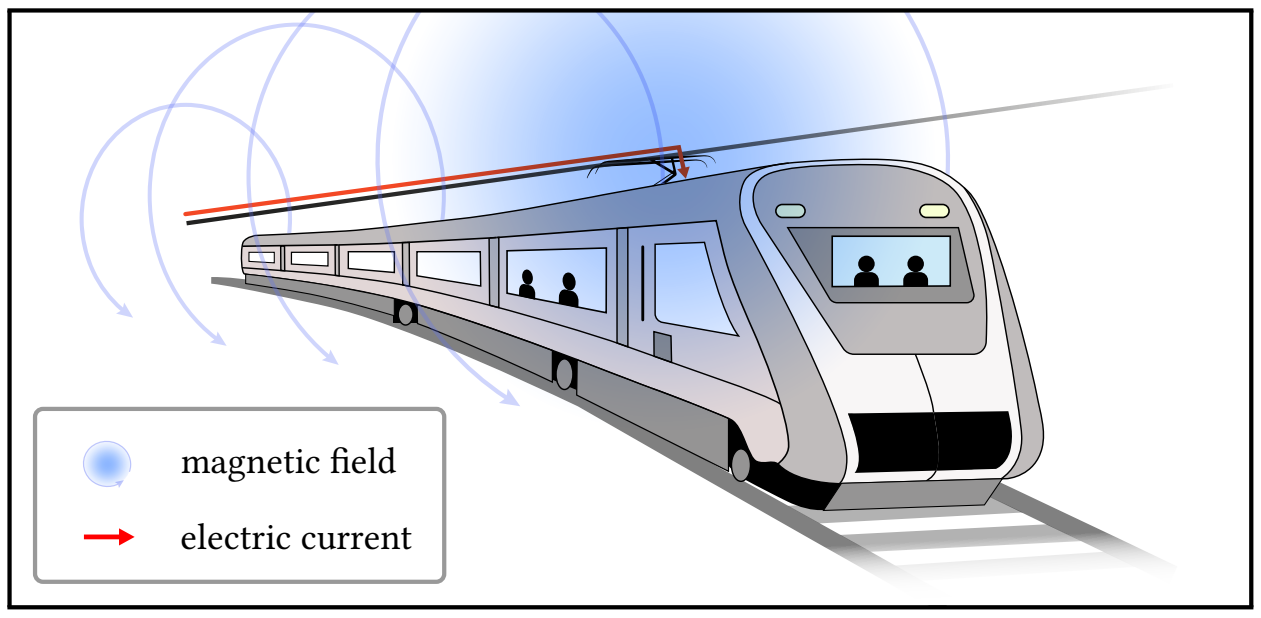
\begin{figure}[t]
    \centering
    \setlength{\fboxrule}{0.8pt}
    \setlength{\fboxsep}{0pt}
    \begin{tikzpicture}
        \node[inner sep=0pt] (img) {
                    {\fbox{\includegraphics[width=0.46\textwidth]{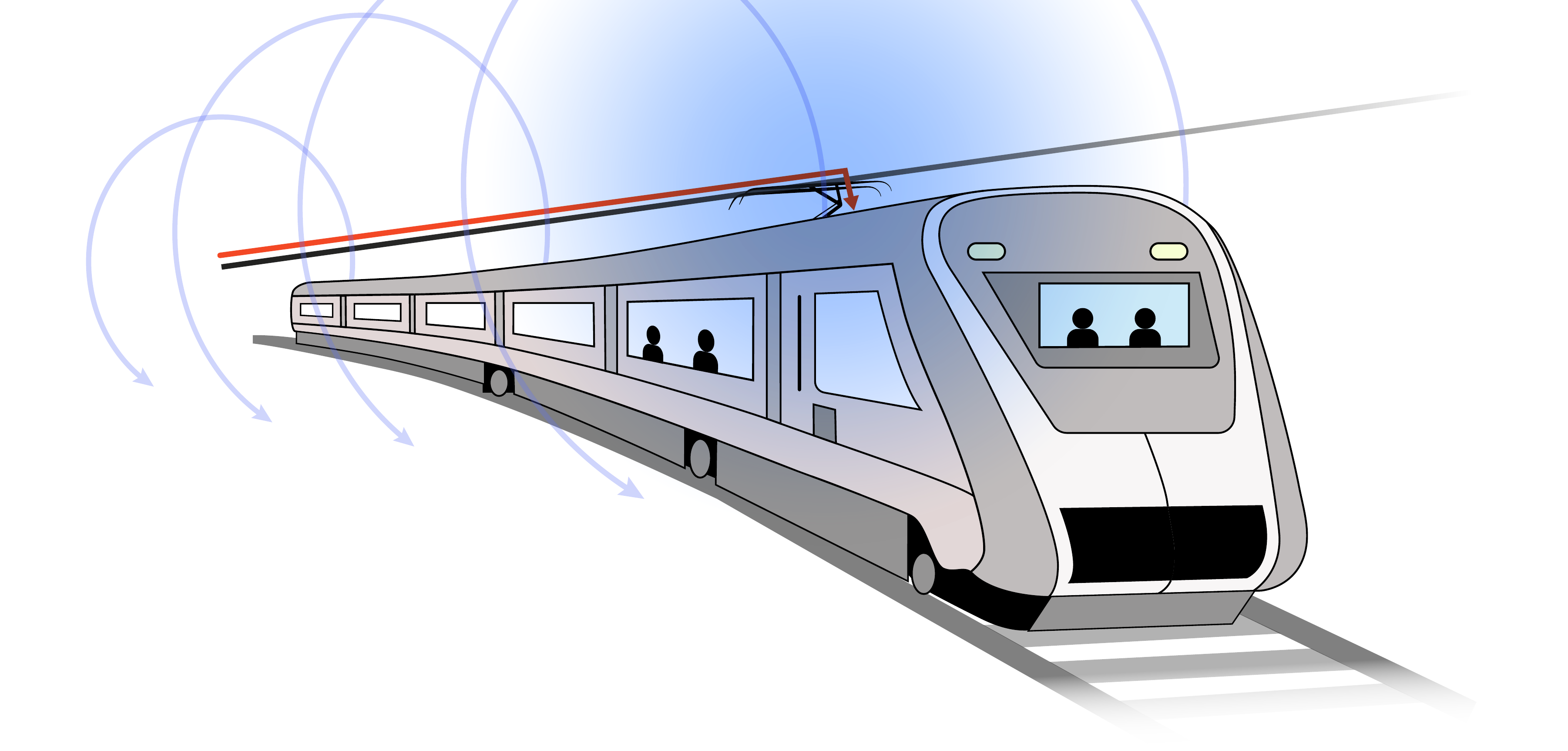}}}
        };
        \node[
            anchor=south west,
            xshift=5pt,
            yshift=5pt,
            rounded corners=1.5pt,
            draw=gray!80,
            line width=0.8pt,
            fill=white,
            fill opacity=0.8,
            text opacity=1,
            inner sep=3pt
        ] at (img.south west) {
            \begin{tikzpicture}[baseline]
                \node[anchor=center] (sym)
                    {\includegraphics[width=0.018\textwidth]{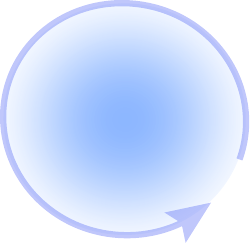}};
                
                \node[anchor=west] at ([xshift=3pt]sym.east)
                    {\fontsize{7}{8.4}\selectfont magnetic field};
                \node[anchor=west] (arrowstart) at ([yshift=-6pt, xshift=-3pt]sym.south west) {};
                \draw[
                    red,
                    line width=0.9pt,
                    -{Triangle[length=3pt,width=3pt]}
                ] (arrowstart) -- ++(0.45,0);
                \node[anchor=west] at ([xshift=15pt]arrowstart.east)
                    {\fontsize{7}{8.4}\selectfont electric current};
            \end{tikzpicture}
        };
    \end{tikzpicture}
    \caption{Train illustration with the magnetic field introduced by the current through the overhead line.}
    \label{fig:train}
\end{figure}

Strong magnetic fields occur in systems with high operating currents. Examples include the electric power grid as well as electric railway and subway systems with overhead lines or conductor rails.
\Cref{fig:train} illustrates a train powered by an overhead line. The traction current flowing through the line produces a magnetic field in the surrounding space, which can be measured by smartphone magnetometers inside the carriage.

\subsection{Electric Trains}

On electrified railways, energy is usually supplied to trains via overhead contact lines and a pantograph mounted on the roof of the train.
The pantograph maintains continuous mechanical contact with the overhead wire while the train is in motion.
In a simplified model the electric current flows from the overhead line through the pantograph to the onboard traction units.
As the current is typically on the order of hundreds of amperes, it produces significant magnetic fields in the vicinity.
The same applies to alternative electrical power transmission schemes, such as the third rail system.

\paragraph{AC and DC traction systems}
Electrified railway networks are widespread across Europe and Asia, while electrification in North America is restricted to selected commuter and intercity corridors. 
Several voltage and frequency standards coexist worldwide. Common \ac{AC} systems are \SI{15}{\kilo\volt} at \SI{16.7}{\hertz} and \SI{25}{\kilo\volt} at \SI{50}{\hertz}. 
In addition, North America has \SI{25}{\hertz} systems, which are typically used for specific railways. 
\ac{DC} systems operate primarily at \SI{1.5}{\kilo\volt} or \SI{3}{\kilo\volt} and are common on subway and tram networks.

For our work, we focus on \SI{16.7}{\hertz} \ac{AC} systems for two reasons.
First, this standard is used in Austria, Germany, Norway, Sweden, and Switzerland, covering most railways in central and northern Europe.
Second, the resulting magnetic field carries a clear \SI{16.7}{\hertz} signature that can be identified directly from the grid frequency.
While \ac{DC} systems also produce magnetic fields, these are static and therefore difficult to distinguish from the Earth's static background.

\paragraph{Magnetic field from traction current}
The traction current flowing through the overhead line produces a magnetic field that can be described using Ampere's law.
In the simplified case of a long straight conductor, the magnetic flux density $B(t)$ is proportional to the electric current $I(t)$:
\begin{equation}
B(t) = \frac{\mu_0}{2\pi r} I(t) = \frac{\mu_0}{2\pi r} \frac{P(t)}{V}
\end{equation}
where $\mu_0$ is the magnetic permeability of free space and $r$ is the distance to the current.
Since the supply voltage $V$ is constant within the system, the current $I$ is proportional to the train's power consumption $P$.
Consequently, if we place a measurement device at a fixed distance from the conductor, the measured magnetic field amplitude directly reflects the train's instantaneous power consumption.
Variations such as acceleration or changes in track gradient therefore produce characteristic time-dependent patterns in the magnetic field.
This even holds on busy tracks: when other trains share the victim's overhead line, the train's movement still produces its own distinctive time-dependent pattern.

\subsection{Smartphone Magnetometers}
\label{sec:smartphone_sensing}

Since our attacks rely on measurements collected from consumer devices, we briefly review the magnetic sensing capabilities of modern smartphones and the constraints that apply to them.
Smartphones are commonly equipped with three-axis \ac{MEMS} magnetometers, which capture the magnetic field vector along all spatial directions.
These sensors are primarily intended for compass functionality and device orientation, but within their accuracy limits they can also be used to record other magnetic signals of interest, including those generated by power systems and electric railway infrastructure.

\paragraph{Sensor access and sampling rates}
The sampling rate of smartphone magnetometers is constrained by both hardware limits and operating system policies.
On Android, as of April 2026, access to magnetometer data does not require explicit runtime permissions.
Since API level 31, however, the sampling rate for all inertial sensors is capped at \SI{200}{\hertz}\footnote{\url{https://developer.android.com/develop/sensors-and-location/sensors/sensors_overview?\#sensors-rate-limiting}}.
This limit exceeds the maximum sampling capabilities of most commercially available magnetometer hardware.
On iOS, currently sensor access is also possible but with an enforced maximum sampling rate of \SI{100}{\hertz}.
Despite these differences, both platforms provide sufficient temporal resolution to capture slowly varying magnetic fields.
This includes signals generated by power systems and electric railway infrastructure, such as the \SI{16.7}{\hertz} frequency used in overhead lines.

\section{Privacy Leaks in Rail Transport}
\label{sec:approach}

Having established the background, we now turn to the privacy leak resulting from magnetic fields in electric trains and show how usable signals can be extracted from the magnetometer.

\subsection{Magnetic Field Signals}
\label{sec:approach_fields}

As discussed in \cref{sec:background}, smartphone magnetometers can measure magnetic fields induced by electric currents.
In particular, the data acquired from the magnetometer consists of three orthogonal components aligned with the device coordinate system.
To obtain an orientation-independent representation of this data, we compute the Euclidean norm of the three axes.
This resulting scalar signal reflects the total magnetic field magnitude and is independent of changes in device orientation. 
This normalization is necessary because the device can be moved arbitrarily during a train ride.

\begin{figure}[hbtp]
\centerline{
\import{figures}{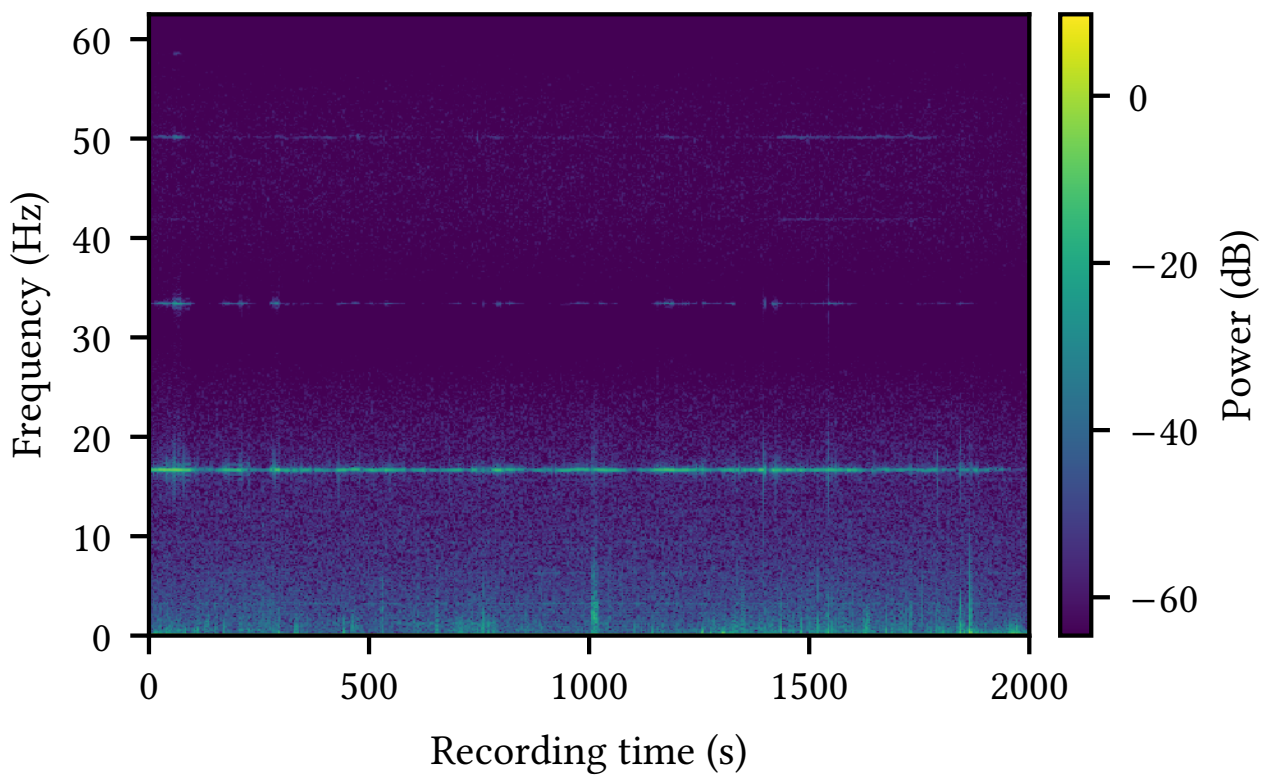}
}
\caption{Spectrogram of the magnetometer of a train ride.}
\label{fig:spectrogram}
\end{figure}

To illustrate the spectral characteristics of the measured magnetic fields, let us consider a spectrogram of the data.
\cref{fig:spectrogram} shows a spectrogram of the magnetic field magnitude for an electric train ride. The Earth's magnetic field appears as a low-frequency component at the bottom of the plot, while a distinct signal at \SI{16.7}{\hertz} corresponds to the electric current supplied via the overhead line. For other railway standards, this signal would appear at the respective grid frequency of \SI{25}{\hertz} or \SI{50}{\hertz}. Although this signal varies over time, it remains clearly visible throughout the entire recording. To work with it further, we apply a notch filter centered at the grid frequency. The filtered signal yields a time-domain trace for investigating changes over time.

\subsection{Magnetic Field Traces}
\label{sec:approach_coloc}

\begin{figure}[b]
    \centerline{
	   \import{figures}{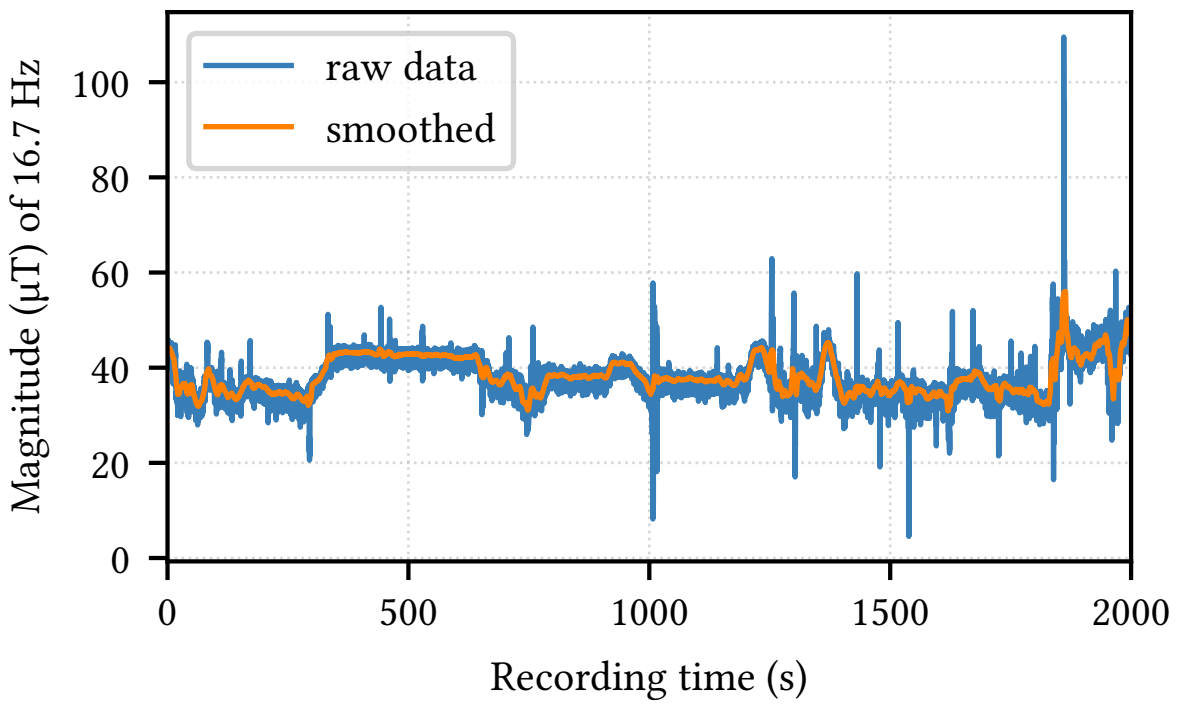}
    } 
    \vspace*{-6pt}
	\caption{Signal of the magnetic field during a train ride.}
	\label{fig:trace}
\end{figure}

\cref{fig:trace} shows the extracted trace from the train ride given in \cref{fig:spectrogram} after filtering at \SI{16.7}{\hertz} and isolating the amplitude of the magnetic field.
We observe distinct features of the train's movement: 
At the start, the signal fluctuates as the train moves out of the station. 
After around 300\,s the train passes a switch between two tracks and follows one track until around 1300\,s, where the next switch occurs.
On the following straight section, the signal remains relatively stable, as train speed is constant and field variations are mainly caused by other trains drawing current from the same overhead line.
After around 1800\,s the train enters a busy rail network, where the signal shows more fluctuation, driven by the high density of other trains on the same overhead line.

To prepare a trace for comparison with others, we apply a high-pass filter to remove the static component and retain only the temporal changes in the signal. Since the amplitude depends on the distance to the overhead line, we normalize each trace by subtracting its mean and dividing by its standard deviation, making it independent of source distance and other attenuation factors. Finally, a smoothing step using a sliding window suppresses short-term noise, yielding a normalized trace that captures the characteristic current pattern of a train ride. This smoothed signal is shown as orange line in \cref{fig:trace}.

\begin{figure}[htb]
    \centerline{
	   \import{figures}{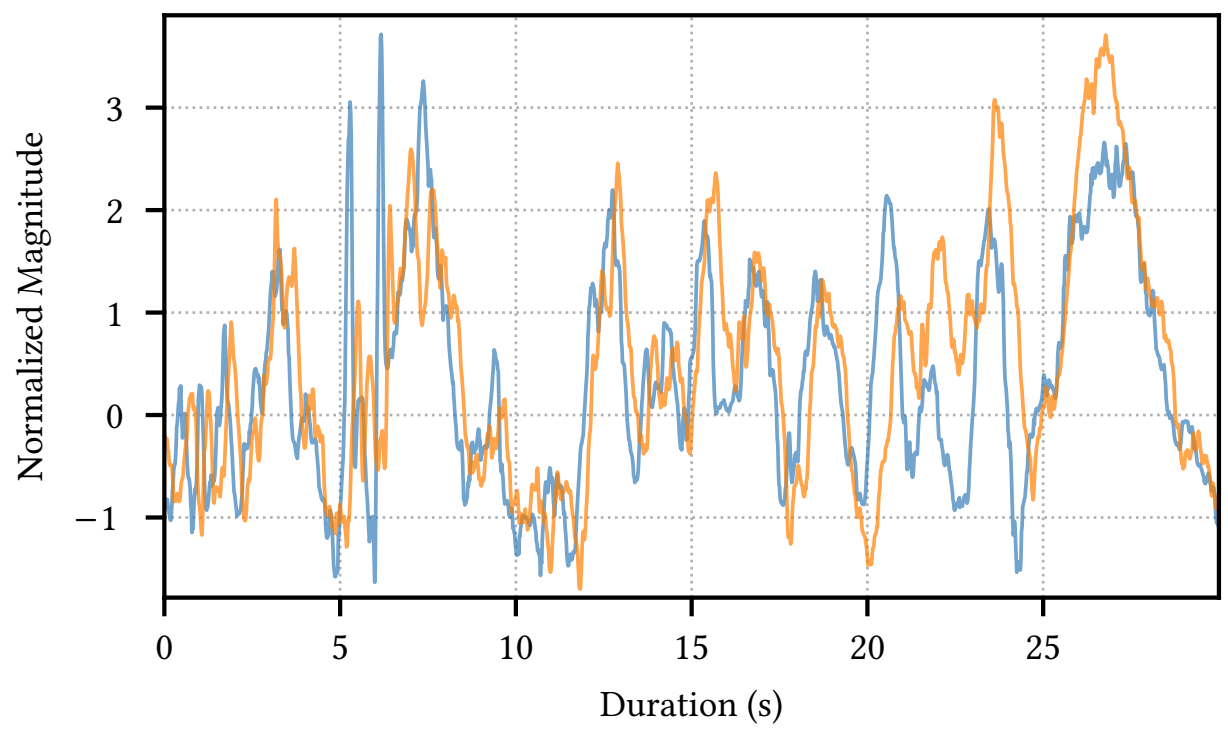}
    }
	\caption{Signals of magnetic fields captured by magnetometers of two smartphones in the same train. }
	\label{fig:trace_coloc_same_car_similar}
\end{figure}

To illustrate the utility of this trace, let us examine the signals of two smartphones placed in different railcars of the same electric train. 
\cref{fig:trace_coloc_same_car_similar} shows the respective traces. 
The recordings exhibit a strong similarity in their magnitude changes, despite minor differences. This shared structure arises because both devices are exposed to the same magnetic field during the train ride. We leverage these shared patterns to determine whether two smartphones are aboard the same train, that is, whether they are \textit{colocated}.

At first glance, the signals shown in \cref{fig:trace,fig:trace_coloc_same_car_similar} may seem to also indicate location-dependent features, such as particular switches on the route or the entering and exiting of rail networks.
However, the trace does not primarily reflect the position of the recording train.
The observed variations in the signal are influenced by the current train as well as all other trains connected to the same overhead line.
Each of them draws or recuperates power depending on changes in speed, driving style, and traffic situation, so a share of the peaks stems from other trains accelerating or braking somewhere on the same feeding section.
These events are time-dependent and vary from ride to ride, which means that the same route does not produce a stable and reproducible trace.
Genuinely location-anchored features, such as section insulators or feeding points, do exist, but they are too sparse to compensate for the dominating temporal component.
Consequently, the trace hinders precise localization or the identification of a particular train line.
At the same time, it provides an accurate fingerprint of temporal events that all passengers of a train observe simultaneously, which is exactly what colocation inference requires.

\section{Attack Scenarios}
\label{sec:attack_scenarios}

Equipped with an understanding of how information leaks from the observed magnetic fields in electric trains, we are now ready to derive privacy attacks and the underlying threat model.

\subsection{Threat Model}
\label{sec:threatmodel}

The attacker's goal is to infer whether a victim is traveling by electric train and which victims are traveling together, without access to any location-related permission.
We consider an attacker who is able to retrieve data recordings from applications running on a user's smartphone.
As the magnetometer is not protected by permissions on the Android or iOS platform, any application can record magnetometer data without raising user suspicion.
This covert monitoring can occur in different scenarios. First, the attacker may compromise a popular smartphone application through a third-party library and observe user movements on a large scale.
For example, an advertisement provider might use the attack to track user habits unnoticed.
Second, the adversary may target specific users through a compromised application tailored to their interests.
For instance, an oppressive regime could monitor user movements via a tampered messaging application.

\paragraph{Attacker Capabilities}
We assume that the attacker-controlled application can access magnetometer data without requesting runtime permissions.  
The application can record sensor data continuously in the background and transmit it to a server if required. 
We do not impose particular constraints on the computing costs for this; however, we later show that our attacks are efficient enough not to raise the user's suspicion.

\medskip
In the following, we propose two attacks that can be combined into a two-stage approach.
In the first stage, the attacker analyzes magnetometer data to detect characteristic frequency components induced by the train's power system, enabling transport mode detection (\cref{sec:approach_tmd}). 
This step can be performed locally on the device and filters the victim population down to users currently on an electric train.
In the second stage, the attacker compares magnetic field traces across devices to identify shared patterns over time.
These shared patterns enable colocation inference between users (\cref{sec:approach_colocation}), even in the absence of explicit location information.
If a reference trace with known location is available, the inferred colocation directly exposes the location of additional users. 

\subsection{Transport Mode Detection}
\label{sec:approach_tmd}

For \ac{TMD}, we do not require a complete amplitude trace of the magnetic field.
Instead, it is sufficient to determine whether the frequency component of the overhead line is present in the magnetometer signal.
As established in \cref{sec:approach}, this frequency is characteristic of electrified railway infrastructure and typically occurs only inside or close to a train.
Detecting its presence therefore allows us to infer whether a victim user is currently traveling by an electric train.

\paragraph{Detection approach.}
To detect the presence of the target frequency, we require a method for identifying specific frequency components in a signal. 
Several techniques are available for this task, such as the \ac{DFT}, \ac{PSD} estimation via Welch's method~\citep{welchUseFastFourier1967}, and the Goertzel algorithm~\citep{goertzelAlgorithmEvaluationFinite1958}. 
Each method offers a different trade-off between implementation complexity, computational cost, frequency resolution, and robustness to noise.
In our scenario, the main decision criterion is the \ac{SNR} at the target frequency. The \ac{SNR} compares the signal at the target frequency to an estimate of the background noise level derived from the surrounding spectrum. If the \ac{SNR} exceeds a defined threshold, we can consider the frequency component \textit{present} and the data is classified as a train ride.

When experimenting with different approaches, we found that the \ac{DFT}-based method consistently achieves high detection performance while remaining computationally lightweight. 
We therefore adopt it as our method of choice. 
In particular, the normalized magnetometer signal is first detrended to remove its static component, then multiplied by a Hann window to reduce spectral leakage at bin boundaries. 
A \ac{DFT} is applied to the windowed signal, and an \ac{SNR} measure around the \SI{16.7}{\hertz} target frequency is computed.
If the \ac{SNR} exceeds a threshold $\tau$, the window is classified as a train ride.

To further stabilize detections over time, we apply a majority vote over a sliding window of consecutive chunks of data, exploiting the fact that spontaneous alternating changes in transport mode are physically unlikely over short periods.

\subsection{Colocation Prediction}
\label{sec:approach_colocation}

For the prediction of colocation, we use the processed magnetometer signal trace described in \Cref{sec:approach}. 
We consider two approaches for realizing the attack: a distance-based method that measures trace similarity directly, and a machine learning-based method that frames colocation as a classification problem.

\paragraph{Distance-based approach.}

A natural fit for inferring colocation of two devices is to quantify the similarity between two traces using a distance metric and apply a threshold: pairs with a distance below the threshold are classified as colocated, while pairs above it are not. 
Simple metrics such as Euclidean distance are limited in this setting, as they are not robust to minor temporal misalignments between traces.
Alignment-aware distances are therefore preferred. 
\Ac{DTW} warps the time axes of both traces to find the non-linear alignment that minimizes the cumulative distance between matched points.
\Ac{DDTW} is a variant that runs \ac{DTW} on the first-order derivatives of the traces, making the comparison sensitive to the shape rather than the level of the signal.
Since our traces are already normalized and reduced to the 16.7 Hz component, the additional shape sensitivity of \ac{DDTW} offers little benefit, while it adds a further step to every comparison.

We therefore adopt \ac{DTW} as the trade-off between detection performance and computational cost.
We extract segments of equal length from both traces, aligned to the same point in time using the device clocks.
This ensures that both traces capture the same events on the train at the same time, maximizing the expected similarity.
The comparison threshold is optimized on a set of recorded magnetometer data.
Besides the threshold, we optimize the number of data points per second retained in the trace and temporal smoothing.
While this approach is conceptually simple, the computational cost scales linearly with trace length and the number of comparisons---a bottleneck for 1:n inference scenarios.

\paragraph{Learning-based approach.}
For a learning-based approach to infer colocation, we draw on recent work in deep learning for time series analysis.
In particular, we employ convolutional neural networks, which are well suited for our task as they can effectively capture local patterns and temporal dependencies~\citep{Yi2025}.
We intentionally employ a lightweight network architecture to ensure that the inference process remains computationally lightweight, fulfilling a crucial requirement for seamless, real-time deployment.

During the training phase, the parameters of the \ac{CNN} are optimized using a comprehensive corpus of prerecorded signal pairs.
We determine them in two steps.
First, we derive the admissible range of each parameter from the physics of the signal, so that the resulting configurations remain able to represent the temporal variations of the traction current.
Second, we search within these ranges for the values that perform best on the training data.
Through this optimization, the network ultimately functions as a powerful, non-linear approximator, capable of directly classifying whether a given pair of recordings indicates device colocation.

Adopting this learning-based methodology yields two fundamental advantages over traditional distance-based comparisons.
First, it completely shifts the burden of feature extraction from rigid mathematical metrics to the network itself.
By learning directly from a training dataset, the model automatically identifies and isolates the specific, subtle signal variations that definitively indicate shared exposure to the same localized magnetic field.
Second, the computational complexity of the evaluation is drastically reduced. Because classifying a pair of signal chunks requires only a single, efficient forward pass through the trained network, the comparison cost scales strictly linearly with the length of the data.
This circumvents the quadratic scaling bottlenecks inherent to techniques like dynamic time warping, making the CNN approach exceptionally scalable for widespread inference tasks.

\section{Evaluation Data}
\label{sec:dataset}

To evaluate both attacks, we collected a comprehensive dataset of smartphone sensor recordings from electric train rides, comprising detailed readings from all inertial sensors. 
While several public datasets exist, such as US-TMD~\cite{carpinetiCustomDualTransportation2018}, Collecty~\cite{erdelicDatasetMultimodalTransport2023}, MobilitApp~\cite{aguilarigartuaMobilitAppMobilitySensor2024}, NOR-TMD~\cite{erdelicDatasetMultimodalTransport2023}, HTC~\cite{yuBigDataSmall2014}, and SHL~\cite{gjoreskiSussexHuaweiLocomotionTransportation2018}, none is directly suitable for our setting, as they do not distinguish between electric and non-electric trains.

\subsection{Data Collection}
\label{sec:dataset_collection}

For data collection, we recruited participants who regularly commute by train and provided them with a short set of instructions for collecting and labeling data.
Recordings were made with the Sensor Logger app \cite{choi_sensorlogger} at the highest sampling rate the device offered, typically 100\,Hz.
We recorded the magnetometer along with the accelerometer, gyroscope, gravity, linear acceleration, orientation and pressure whenever they were available.
This broader sensor set was deliberately captured to make the data useful beyond the present work.

To create labels, participants noted each stop and departure together with the station name, the train line number, and the train type.
The resulting labels divide each recording into trips and into station-to-station segments, with the train type allowing us to separate electrified from non-electrified trains.
We placed no restrictions on smartphone usage during the recordings.
Participants used their devices as they normally would, including phone calls, messaging, and web browsing, as well as moving the device between pocket, hand, and table.
The dataset therefore reflects realistic in-the-wild conditions rather than a controlled placement protocol.

\subsection{Dataset Composition}
\label{sec:dataset_composition}
The collected data contains \TrainTrackRecordingDuration\,hours of labeled recordings.
The data is organized hierarchically: a \emph{journey} is a continuous recording session of a participant's travel, a \emph{trip} is a single train ride between two changes within a journey, and a \emph{segment} is a station-to-station segment within a trip.
The released dataset comprises \TrainTrackJourneys journeys, \TrainTrackTrips trips, and \TrainTrackSegments segments, covering \TrainTrackStations unique stations.
There are in total \ColocDatasetAllSegmentsMax segments with at least two colocated recordings in the same train. 

\begin{table}[thbp]
\centering \small
\caption{
Distribution of train types within the collected data. \# is the number of segments for the train type.}
\label{tab:dataset_overview}
\begin{tabular}{@{}lrrrrrr@{}}
\toprule
\multicolumn{1}{c|}{\multirow{2}{*}{\textbf{Type}}} & \multicolumn{2}{c|}{\textbf{All}}                                     & \multicolumn{4}{c}{\textbf{16.7\, Hz electric}}                                                                                                                                   \\
\multicolumn{1}{c|}{}                               & \multicolumn{1}{c}{\textbf{hours}} & \multicolumn{1}{c|}{\textbf{\#}} & \multicolumn{2}{c}{\textbf{hours}}                                                         & \multicolumn{2}{c}{\textbf{\#}}                                                      \\ \midrule
Long distance                                       & \TrainTrackLongDistanceDuration    & \TrainTrackLongDistanceSegments  & \TrainTrackLongDistanceElectricYesDuration & \TrainTrackLongDistanceElectricYesDurationPct & \TrainTrackLongDistanceElectricYesCount & \TrainTrackLongDistanceElectricYesCountPct \\
Regional                                            & \TrainTrackRegionalDuration        & \TrainTrackRegionalSegments      & \TrainTrackRegionalElectricYesDuration     & \TrainTrackRegionalElectricYesDurationPct     & \TrainTrackRegionalElectricYesCount     & \TrainTrackRegionalElectricYesCountPct     \\
Urban                                               & \TrainTrackUrbanDuration           & \TrainTrackUrbanSegments         & \TrainTrackUrbanElectricYesDuration        & \TrainTrackUrbanElectricYesDurationPct        & \TrainTrackUrbanElectricYesCount        & \TrainTrackUrbanElectricYesCountPct        \\
Metro                                               & \TrainTrackMetroDuration           & \TrainTrackMetroSegments         & \TrainTrackMetroElectricYesDuration        & \TrainTrackMetroElectricYesDurationPct        & \TrainTrackMetroElectricYesCount        & \TrainTrackMetroElectricYesCountPct        \\
Tram                                                & \TrainTrackTramDuration            & \TrainTrackTramSegments          & \TrainTrackTramElectricYesDuration         & \TrainTrackTramElectricYesDurationPct         & \TrainTrackTramElectricYesCount         & \TrainTrackTramElectricYesCountPct         \\ \midrule
Total                                               & \TrainTrackRecordingDuration       & \TrainTrackSegments              & \TrainTrackElectricYesDuration             & \TrainTrackElectricYesDurationPct             & \TrainTrackElectricYesCount             & \TrainTrackElectricYesCountPct             \\ \bottomrule
\end{tabular}
\end{table}

\cref{tab:dataset_overview} shows the distribution of train types, both in recording time and in number of station-to-station segments.
Long-distance trips dominate the dataset by recording time, while urban transport contributes a comparable number of shorter segments.
The \SI{16.7}{\hertz} signature exploited by our attacks is present mainly in regional and long-distance trains; metros and trams typically use other power supply systems.

This dataset is comparable in size to other public datasets, yet it contains, to the best of our knowledge, the largest number of train rides available to date.
Moreover, it is the only one collected with electrification labels.

\subsection{Dataset Availability and Privacy}
\label{sec:dataset_release}

The original collected dataset comprises 137\,GB and is available at: \url{https://doi.org/10.5281/zenodo.22206115}.

To protect participant privacy, the dataset is anonymized. Station names are replaced with random identifiers, with the same physical station mapped consistently across journeys. %
Absolute timestamps are removed and replaced with times relative to the start of each journey; the relative timing between journeys recorded simultaneously is preserved, so that colocated trips remain identifiable.
Finally, in line with privacy regulations, all participants provided informed consent for the release of the dataset. 
Our research institution does not operate an Institutional Review Board (IRB), as is common at European institutions; we therefore followed the principles of the Menlo Report, as detailed in \Cref{sec:ethics}.

\section{Evaluating Transport Mode Detection}
\label{sec:tmd}

We now evaluate our first attack: \acf{TMD} in electric trains. In particular, after describing the experimental setup, we compare our approach against existing baselines and report its detection performance.

\subsection{Experimental Setup}
\label{sec:tmd_setup}

To evaluate our approach, we derive an evaluation dataset from our recordings; the preprocessing pipeline and implementation details are described below.

\paragraph{Data preprocessing}

For this experiment, we consider all electric train rides operating on a 16.7\,Hz grid frequency from our dataset. This includes all long-distance and regional trips. As negative samples, we additionally consider recordings collected during daily activities in which participants carried their smartphones outside of train rides.
For these recordings, we leverage the Google Activity Recognition API\footnote{\url{https://developers.google.com/location-context/activity-recognition}} to infer the current activity, including classes such as Stationary, Walking, Cycling and Automotive.
An overview of the data distribution is shown in \cref{tab:tmd_activity}.
We perform the split between training and testing data at the journey level, assigning \PctPrint{30} of the journeys to the test set.

\newcommand{\TmdNoTrainActivityCyclingTime}{\numprint{21983}\,s\xspace}
\newcommand{\TmdNoTrainActivityStationaryTime}{\numprint{107925}\,s\xspace}
\newcommand{\TmdNoTrainActivityUnknownTime}{\numprint{89935}\,s\xspace}
\newcommand{\TmdNoTrainActivityWalkingTime}{\numprint{33048}\,s\xspace}
\newcommand{\TmdNoTrainActivityAutomotiveTime}{\numprint{29598}\,s\xspace}
\newcommand{\TmdNoTrainActivityTiltingTime}{\numprint{38}\,s\xspace}
\newcommand{\TmdNoTrainActivityTotalTime}{\numprint{282527}\,s\xspace}

\begin{table}[thbp]
\centering \small
\caption{
Overview of the data for transport mode detection. Activities are from Google's Activity Recognition API.}
\label{tab:tmd_activity} 
\begin{tabular}{@{}lrlr@{}}
\toprule
\textbf{Activity} & \textbf{Time}                     & \textbf{Activity} & \textbf{Time}                     \\ \midrule
Stationary        & \TmdNoTrainActivityStationaryTime & Automotive        & \TmdNoTrainActivityAutomotiveTime \\
Walking           & \TmdNoTrainActivityWalkingTime    & Tilting           & \TmdNoTrainActivityTiltingTime    \\
Cycling           & \TmdNoTrainActivityCyclingTime    & Unknown           & \TmdNoTrainActivityUnknownTime    \\ \midrule
No Train          & \TmdNoTrainActivityTotalTime      & Train             & \TmdTrainActivityTime             \\ \bottomrule
\end{tabular}
\end{table}

\paragraph{Implementation details}
For signal processing an \ac{FFT} is applied to the windowed signal, yielding a single spectrum over the full observation window with frequency resolution $\Delta f = f_s / N$, where $f_s$ is the sampling rate and $N$ is the number of samples. 
We then identify the peak amplitude within a narrow band of $\pm\delta_s$~\si{\hertz} around the target frequency of \SI{16.7}{\hertz} and compute the \ac{SNR} against the estimated noise floor. 
If the \ac{SNR} exceeds a threshold $\tau$, the window is classified as a train ride. Both $\delta_s$ and $\tau$ are treated as hyperparameters and optimized on the training set.

These free parameters are optimized with Optuna~\cite{akiba2019optuna}, a black-box hyperparameter optimization framework.
The tuned parameters are the width of the signal band, $\delta_s$, the \ac{SNR} threshold $\tau$, and the width of the noise band $\delta_n$.
The search ranges for these parameters are derived from the physics of the signal: the signal band $\delta_s$ has to be wide enough for the traction frequency and the resolution $\Delta f$ of the transform, yet narrow enough to exclude neighboring spectral content, and the noise band $\delta_n$ has to cover enough bins for a stable noise floor estimate without reaching into the signal band.
For noise floor estimation, we evaluate three strategies: the median of non-target bins, linear interpolation between neighboring bands, and a $1/f$ spline fit.
We run the search for \numprint{5000} trials, and each trial repeats the evaluation across 5 random journey-level splits of the training and test set to obtain a robust estimate of the objective.
For each resulting detector, we then evaluate majority-vote aggregation over chunk lengths of $[1, 3, 6, 9, 12]$ seconds.
We use the \acf{MCC} as our primary evaluation metric, as it is robust to class imbalance, and additionally report the F1 score for comparability with prior work on \ac{TMD}.

\subsection{Baseline Approaches}
\label{sec:tmd_baselines}
Over the past decades, \ac{TMD} has been extensively studied, as discussed in detail in \cref{sec:related_work}.
In \cref{tab:tmd_related_work}, we summarize all studies that, to the best of our knowledge, rely exclusively on permissionless smartphone sensors and include at least one class corresponding to train journeys.

\begin{table}[thbp]
\centering \small
\caption{
Related approaches and datasets. Sensors are Accelerometer (A), Gravity (G), Gyroscope (Y), Magnetometer (M) and Pressure (P). C is the number of \ac{TMD} classes.
\label{tab:tmd_related_work}
}
\begin{tabular}{@{}llcrl@{}}
\toprule
\textbf{Study}                                         & \multicolumn{1}{c}{\textbf{Year}} & \textbf{Dataset} & \multicolumn{1}{c}{\textbf{C}} & \multicolumn{1}{c}{\textbf{Sensors}} \\ \midrule
\citet{hemminkiAccelerometerbasedTransportationMode2013} & 2013                              &                  & 7                              & A                                    \\
\citet{iskanderovBreakingLimitsTransportation2020}       & 2020                              & HTC              & 13                             & A, Y, M                              \\
\citet{wangTransportationModeDetection2022}              & 2022                              & SHL, HTC         & 8                              & A, Y, M, P                           \\
\citet{wangDFTransDualFrequency2022}                     & 2022                              & SHL, HTC         & 8                              & A, Y, M, P                           \\
\citet{taherinavidAutomaticTransportationMode2024}       & 2024                              & HTC              & 8                              & A, G, Y, M                           \\
\citet{xuAutomaticClassificationTransportation2024}      & 2024                              & HTC              & 8                              & A, G, Y, M                           \\
\citet{ibanezMobilitAppDeepLearningBased2025}            & 2025                              & MobilitApp       & 11                             & A, Y, M                              \\
\citet{skrettingFeatureImportanceNovel2025}              & 2025                              & NOR-TMD          & 8                              & A, G, Y, M                           \\ \midrule
Our study                                              & 2026                              & Ours             & 2                              & M                                    \\ \bottomrule
\end{tabular}
\end{table}

Previous work on \ac{TMD} typically addresses multi-class classification, where the goal is to distinguish between several transport modes.
This setting is inherently more challenging than our binary classification task.
Nevertheless, these approaches provide a useful basis for comparison.
To enable comparability, we extract the reported F1 and \ac{MCC} scores from the respective studies.
In cases where multiple train-related classes are present, such as \emph{Train} and \emph{High-Speed-Train}, we merge them into a single class.
Although the approaches in \cref{tab:tmd_comparisson} are evaluated on different datasets under varying conditions, they serve as a useful reference point for contextualizing the performance of existing methods.

\subsection{Results}
\label{sec:tmd_results}

\Cref{tab:tmd_comparisson} shows the performance of our attack in comparison to results reported by prior work. 
Existing approaches for detecting train rides achieve F1 scores ranging from 0.70 to 0.93 on their respective datasets. 
All studies except \citet{ibanezMobilitAppDeepLearningBased2025} use \SI{10}{\second} chunks for classification; the latter does not report sufficient information to compute the \ac{MCC} score.
To reproduce the results of \citet{skrettingFeatureImportanceNovel2025} we reimplemented their approach, outlined in \cref{app:nor_tmd}.
Despite relying on a single sensor and a single spectral feature, our \ac{TMD} attack achieves an F1 score of \TmdResultBestFOneChunk and an \ac{MCC} of \TmdResultBestMccOneChunk, placing it on par with state-of-the-art approaches. 
Applying a majority vote further improves performance to an F1 score of \TmdResultBestFOverall and an \ac{MCC} of \TmdResultBestMccOverall, outperforming prior work in our evaluation setting.

\begin{table}[thbp]
\centering \small
\caption{
Performance of \ac{TMD} for our attack and related work. Performance numbers are only shown for the ``train'' class of related approaches.%
\label{tab:tmd_comparisson}
}
\begin{tabular}{@{}lcrcc@{}}
\toprule
\multicolumn{1}{c}{\textbf{Study}}                     & \textbf{Dataset} & \multicolumn{1}{c}{\textbf{F1}} & \multicolumn{1}{c}{\textbf{MCC}} \\ \midrule
\citet{hemminkiAccelerometerbasedTransportationMode2013} &                                                                          & 0.70                            & 0.68                             \\
\citet{iskanderovBreakingLimitsTransportation2020}       & HTC                                                                    & 0.83                            & 0.79                             \\
\citet{wangTransportationModeDetection2022}              & SHL                                                                     & 0.74                            & 0.72                             \\
\citet{wangTransportationModeDetection2022}              & HTC                                                                      & 0.79                            & 0.77                             \\
\citet{wangDFTransDualFrequency2022}                     & SHL                                                                     & 0.74                            & 0.73                             \\
\citet{wangDFTransDualFrequency2022}                     & HTC                                                                     & 0.83                            & 0.82                             \\
\citet{ibanezMobilitAppDeepLearningBased2025}            & MobilitApp                                                              & 0.66                   & -                       \\
\citet{skrettingFeatureImportanceNovel2025}              & NOR-TMD                                                                 & 0.93                   & 0.93                    \\ \midrule
Our attack w/o majority vote                                              & Our dataset                                                              & \TmdResultBestFOneChunk                            & \TmdResultBestMccOneChunk                             \\
Our attack w/ majority vote                                              & Our dataset                                                               & \textbf{\TmdResultBestFOverall}                           & \textbf{\TmdResultBestMccOverall}                             \\\bottomrule
\end{tabular}
\end{table}

\paragraph{Effect of data cardinality and granularity}

\begin{figure}[htbp]
    \centerline{
       \import{figures}{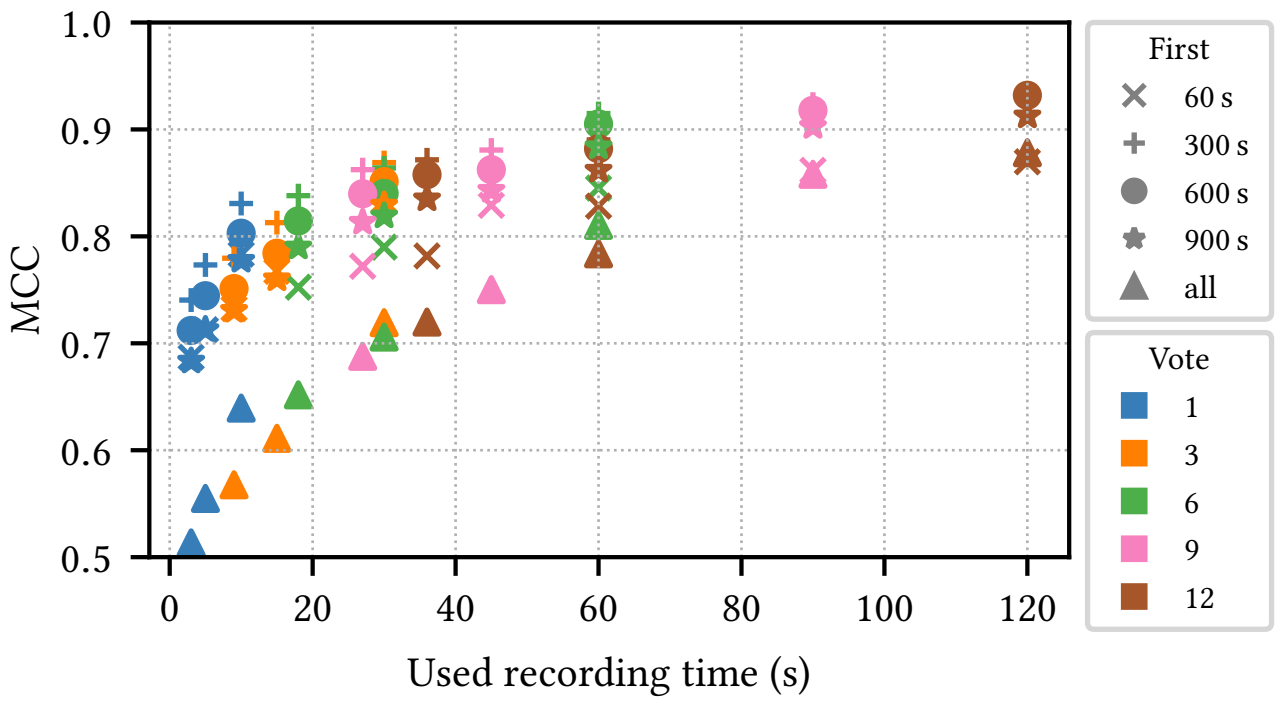}
    }
    \caption{Influence of recording time on \ac{TMD} performance with and without majority vote.}
    \label{fig:tmd_results_recording_time}
\end{figure}

\Cref{fig:tmd_results_recording_time} shows how the total amount of data used for detection influences the \ac{MCC} score of our attack.
The x-axis represents the total data used, computed as the product of the number of chunks in the majority vote and the chunk duration.
We observe that more data consistently improves detection across all post-departure windows.
Whether this data is spread across a longer \ac{DFT} window or across more majority-vote chunks has little influence on the result.
With as little as \SI{60}{\second} of data, our detector already achieves an \ac{MCC} of \TmdResultBestMccFirstThreeHundredDTenWSix.

\paragraph{Effect of the post-departure window}
The effect of recording length on performance is shown in \cref{tab:tmd_recording_length}.
Across all majority-vote lengths, performance is highest when restricting the analysis to the start of the train ride, and degrades as longer segments than \SI{600}{\second} are included.
The reason is physical: trains draw large traction currents during acceleration, which produce a strong \SI{16.7}{\hertz} component.
Including later, low-signal segments dilutes the feature and reduces performance.
Using all data further introduces a larger class imbalance, with fewer non-train chunks, which explains the gap between F1 and \ac{MCC} in \cref{tab:tmd_comparisson}.
This is why we adopt \ac{MCC} as the primary metric throughout.

\begin{table}[htbp]
  \centering \small
  \caption{Influence of post-departure window on \ac{TMD} performance with majority voting as MCC score.}
  \label{tab:tmd_recording_length}
  \begin{tabular}{rrrrrr}
    \toprule
    \multirow{2}{*}{\textbf{First (s)}}
      & \multicolumn{5}{c}{\textbf{Chunks for majority voting}} \\
    \cmidrule(lr){2-6}
      & 1 & 3 & 6 & 9 & 12 \\
    \midrule
    60
      & \TmdResultBestMccFirstSixtyDTenWOne
      & \TmdResultBestMccFirstSixtyDTenWThree
      & \TmdResultBestMccFirstSixtyDTenWSix
      & \TmdResultBestMccFirstSixtyDTenWNine
      & \TmdResultBestMccFirstSixtyDTenWTwelve \\
    300
      & \TmdResultBestMccFirstThreeHundredDTenWOne
      & \TmdResultBestMccFirstThreeHundredDTenWThree
      & \TmdResultBestMccFirstThreeHundredDTenWSix
      & \TmdResultBestMccFirstThreeHundredDTenWNine
      & \textbf{\TmdResultBestMccFirstThreeHundredDTenWTwelve} \\
    600
      & \TmdResultBestMccFirstSixHundredDTenWOne
      & \TmdResultBestMccFirstSixHundredDTenWThree
      & \TmdResultBestMccFirstSixHundredDTenWSix
      & \TmdResultBestMccFirstSixHundredDTenWNine
      & \textbf{\TmdResultBestMccFirstSixHundredDTenWTwelve} \\
    900
      & \TmdResultBestMccFirstNineHundredDTenWOne
      & \TmdResultBestMccFirstNineHundredDTenWThree
      & \TmdResultBestMccFirstNineHundredDTenWSix
      & \TmdResultBestMccFirstNineHundredDTenWNine
      & \TmdResultBestMccFirstNineHundredDTenWTwelve \\
    All
      & \TmdResultBestMccFirstAllDTenWOne
      & \TmdResultBestMccFirstAllDTenWThree
      & \TmdResultBestMccFirstAllDTenWSix
      & \TmdResultBestMccFirstAllDTenWNine
      & \TmdResultBestMccFirstAllDTenWTwelve \\
    \bottomrule
  \end{tabular}
\end{table}

\paragraph{Effect of majority voting}
\label{sec:tmd_majority}
The majority vote provides a strong improvement at almost no additional cost.
For \SI{10}{\second} chunks within the first \SI{300}{\second} after departure, a single decision yields \TmdResultBestMccFirstThreeHundredDTenWOne, three votes yield \TmdResultBestMccFirstThreeHundredDTenWThree, and twelve votes yield \TmdResultBestMccFirstThreeHundredDTenWTwelve.
The same trend holds in the \SI{60}{\second} window, where the \ac{MCC} rises from \TmdResultBestMccFirstSixtyDTenWOne to \TmdResultBestMccFirstSixtyDTenWTwelve across the same vote lengths.
Transport-mode changes occur on the order of minutes, so aggregating consecutive decisions removes spurious classifications without loss of temporal precision.

\subsection{Summary}
\label{sec:tmd_summary}
Our results show that electric train rides can be detected reliably from the smartphone magnetometer alone.
The feature that carries the decision is the presence of the \SI{16.7}{\hertz} magnetic field generated by electrified railway infrastructure.
Detecting this spectral component requires no learned features and no multi-sensor fusion.
Our detector operates directly on the \ac{DFT} and exposes only a few parameters.

The simplicity of the approach is matched by its data efficiency.
A single \SI{10}{\second} chunk shortly after departure already reaches an \ac{MCC} of \TmdResultBestMccOneChunk, and a short majority vote over \SI{120}{\second} raises this to \TmdResultBestMccOverall.
This places our detector in the same performance range as prior multi-sensor pipelines, while keeping the computational footprint small enough for continuous background operation on a mobile device.
The detector can also serve as a component within existing TMD systems.
Many of these systems rely on machine-learning models trained on multiple sensors and struggle to separate trains from other smooth transport modes such as cars or buses.
A dedicated indicator for the \SI{16.7}{\hertz} component can be added to such pipelines as an auxiliary feature, where it provides a physically grounded signal that is hard to confuse with non-rail transport.

\section{Evaluation of Colocation Prediction}
\label{sec:coloc}

We continue to evaluate our attack for inferring the colocation of devices introduced in \cref{sec:approach_coloc}.
After describing the experimental setup, we introduce our baseline and then evaluate the distance-based and learning-based variants of our approach.
For both variants, we report detection performance as well as the amount of data required before a reliable decision can be made.

\subsection{Experimental Setup}
\label{sec:coloc_task}
We evaluate the prediction of colocation on traces derived from our dataset, and describe the preprocessing, evaluation setup, and chunk aggregation below.

\paragraph{Data preprocessing}
From the recordings dataset described in \cref{sec:dataset}, we extract all trips that occur at least twice to form colocated pairs.
For each trip, we derive the \SI{16.7}{\hertz} trace from the full sensor recording following the approach in \cref{sec:approach}.
The resulting evaluation dataset consists of up to \ColocDatasetAllSegmentsFirstTen trip segments with colocation.
We preprocess each trace as described in \cref{sec:approach_coloc}, including normalization and downsampling to a fixed value of data points per second.
This downsampling ensures consistent time alignment across all traces, reduces the input length to a constant size, and lowers the computational cost of the \ac{DTW} distance calculation.

\paragraph{Evaluation Setup}
The following setup applies to both the distance-based and the learning-based variants of our attack.
We evaluate our approach using time windows from the train ride start of 60, 300, 600, and 900 seconds.
For the learning-based variant, we additionally evaluate the full recordings.
We omit this setting for the distance-based variant, due to the high computational costs and small interest.
Our primary goal is to minimize the amount of data required for reliable inference, reducing both the time an attacker must wait before making a decision and the computational resources needed.
In addition, we investigate smoothing the data using a sliding window to reduce noise and suppress short-term fluctuations in the signal.

To study the impact of different train types, we additionally construct separate datasets for short-distance and long-distance trains.
We perform the train and test data split at the journey level, assigning 30\,\% of the journeys to the test set.
Splitting at the journey level ensures that the traces from a journey are not split across training and evaluation data.
To assess the stability of our results, we repeat each evaluation across 10 independent splits and report the mean.
As evaluation metrics, we use \ac{MCC} as our primary metric, since it is robust to the class imbalance in our dataset, and additionally report F1 for comparability with prior work.

\paragraph{Chunk Aggregation}
The two variants process the chunks differently, which we briefly outline here and in detail in \cref{sec:coloc_dtw,sec:coloc_ml}.
The distance-based variant concatenates all chunks within the post-departure window into a single trace and computes one distance value per trace pair.
No per-chunk decision and no majority vote are applied.
The learning-based variant operates on individual chunks during training and inference, producing a per-chunk decision.
These per-chunk decisions are then aggregated via majority vote to obtain the final trace-level prediction.

\subsection{Baseline}
\label{sec:coloc_baseline}

In contrast to \ac{TMD}, only a limited number of studies address colocation prediction, most of which originate from pandemic tracing research, as discussed in \cref{sec:related_work}.
The work most closely related and comparable to ours is \citet{nguyenColocationEpidemicTracking2017}.
In that study, the authors use the magnitude of the magnetic field and the \ac{DDTW} distance to infer colocation between devices.

Despite this conceptual similarity, there are important differences between their evaluation setting and ours.
In our evaluation, we always compare traces of equal length, which removes trace length as a discriminative feature and makes the task substantially harder.
When traces of different length are compared, the length itself already provides a strong cue, as colocated devices typically share the same recording duration.
By fixing the length across all pairs, the decision has to rely entirely on the shape of the signal rather than on trivial metadata.

\paragraph{Implementation Details}

The baseline approach is conceptually simple: it operates on the magnitude of the magnetic field and applies \ac{DDTW} as the distance measure.
We therefore just calculated the Euclidean norm of the sensor's three axes as described in \cref{sec:approach_fields}.
The resulting traces are then directly applicable to the \ac{DDTW} and we calculated the distance to all combinations of traces.
On the training set we calculate a threshold, which then is used to evaluate the colocation on the testing set. 

\paragraph{Results}
We expect a noticeably lower performance than reported in the original study.
The reason is twofold.
First, the number of trace pairs in our evaluation grows quadratically with the size of the dataset, increasing the chance of spurious matches.
Second, trace length, which is highly discriminative in their setting, is controlled for in ours.
Under these conditions, the baseline has to rely solely on the signal shape, which is a considerably more demanding task.
We report the resulting performance in \cref{sec:coloc_summary} alongside our own approaches and a more detailed ablation study in \cref{app:coloc_baseline}.

\subsection{Distance-based Approach}
\label{sec:coloc_dtw}

As first approach to colocation prediction, we use a distance-based method that quantifies the similarity between two traces and applies a threshold to decide whether they originate from colocated devices.

\paragraph{Implementation Details}
For each trace pair, we used two continuous traces, as described in \cref{sec:coloc_task}.
We then compute the \ac{DTW} distance between these traces.
We adopt a one-versus-all strategy in which each trace is compared against all other traces.
A trace pair is classified as colocated if its distance falls below a decision threshold, and as non-colocated otherwise.
The threshold is optimized on the training split together with the length of the post-departure window, the sampling rate of the underlying signal, and the temporal smoothing.
The optimized threshold is then applied to the held-out test split to identify the smallest amount of data for a reliable colocation prediction.

\paragraph{Results}

\begin{table}[hb]
  \centering
  \caption{Distance-based colocation prediction performance as MCC score depending on data points per second and the length of the compared recording from the start.}
  \small
  \label{tab:coloc_distance_results}
  \begin{tabular}{rcccc}
    \toprule
    \multirow{2}{*}{\textbf{First (s)}}
      & \multicolumn{4}{c}{\textbf{Data points per second}} \\
    \cmidrule(lr){2-5}
      & 10 & 20  & 40 & 60 \\
    \midrule
    60
      & \ColocDistanceDTWROneMCCFirstSixtySRTen
      & \ColocDistanceDTWROneMCCFirstSixtySRTwenty
      & \ColocDistanceDTWROneMCCFirstSixtySRFourty
      & \ColocDistanceDTWROneMCCFirstSixtySRSixty \\
    300
      & \ColocDistanceDTWROneMCCFirstThreeHundredSRTen
      & \ColocDistanceDTWROneMCCFirstThreeHundredSRTwenty
      & \ColocDistanceDTWROneMCCFirstThreeHundredSRFourty
      & \ColocDistanceDTWROneMCCFirstThreeHundredSRSixty \\
    600
      & \ColocDistanceDTWROneMCCFirstSixHundredSRTen
      & \ColocDistanceDTWROneMCCFirstSixHundredSRTwenty
      & \ColocDistanceDTWROneMCCFirstSixHundredSRFourty
      & \ColocDistanceDTWROneMCCFirstSixHundredSRSixty \\
    900
      & \ColocDistanceDTWROneMCCFirstNineHundredSRTen
      & \ColocDistanceDTWROneMCCFirstNineHundredSRTwenty
      & \ColocDistanceDTWROneMCCFirstNineHundredSRFourty
      & \textbf{\ColocDistanceDTWROneMCCFirstNineHundredSRSixty} \\
    \bottomrule
  \end{tabular}
\end{table}

\Cref{tab:coloc_distance_results} shows the \ac{MCC} score of the distance-based approach using \ac{DTW}.
The best performance is achieved at a recording length of \SI{900}{\second} from the train ride start, 60 data points per second, and a sliding window of 150 samples for temporal smoothing, yielding an \ac{MCC} score of \ColocDistanceDTWBestMCC.
Shorter recordings reduce the amount of signal available for distance computation and lead to a noticeable drop in performance.
At \SI{60}{\second}, the \ac{MCC} decreases to 0.17, and even shorter recordings degrade further.
Longer recordings have little influence on the \ac{MCC}, except at 60 data points per second.
A rate of 60 data points per second consistently outperforms lower rates, indicating that finer temporal resolution preserves discriminative details in the \SI{16.7}{\hertz} trace.
However, the higher sampling rate  produces a larger standard deviation across runs, indicating that the approach becomes less stable and reliable.
Beyond the moderate detection performance, the distance-based approach is computationally more expensive, which we investigate in \cref{sec:runtime_analysis}.

\subsection{Learning-based Approach}
\label{sec:coloc_ml}
For our learning-based approach, as introduced in \cref{sec:approach_colocation}, we utilize a Siamese 1D \ac{CNN} with convolutional blocks consisting of a 1D convolutional layer, batch normalization, a ReLU activation, and a max pooling layer.
Both recordings pass through the same convolutional blocks, so their feature sequences stay time-aligned and can be compared at each time step via their absolute difference and element-wise product.
A further convolutional block encodes this comparison, followed by global average and max pooling, allowing the network to learn cross-trace features while encoding both signals identically.
The pooled features are followed by fully-connected layers, which perform the actual binary classification.                                                                                                                                     

\paragraph{Implementation details}
\begin{figure}[htbp]
    \import{figures}{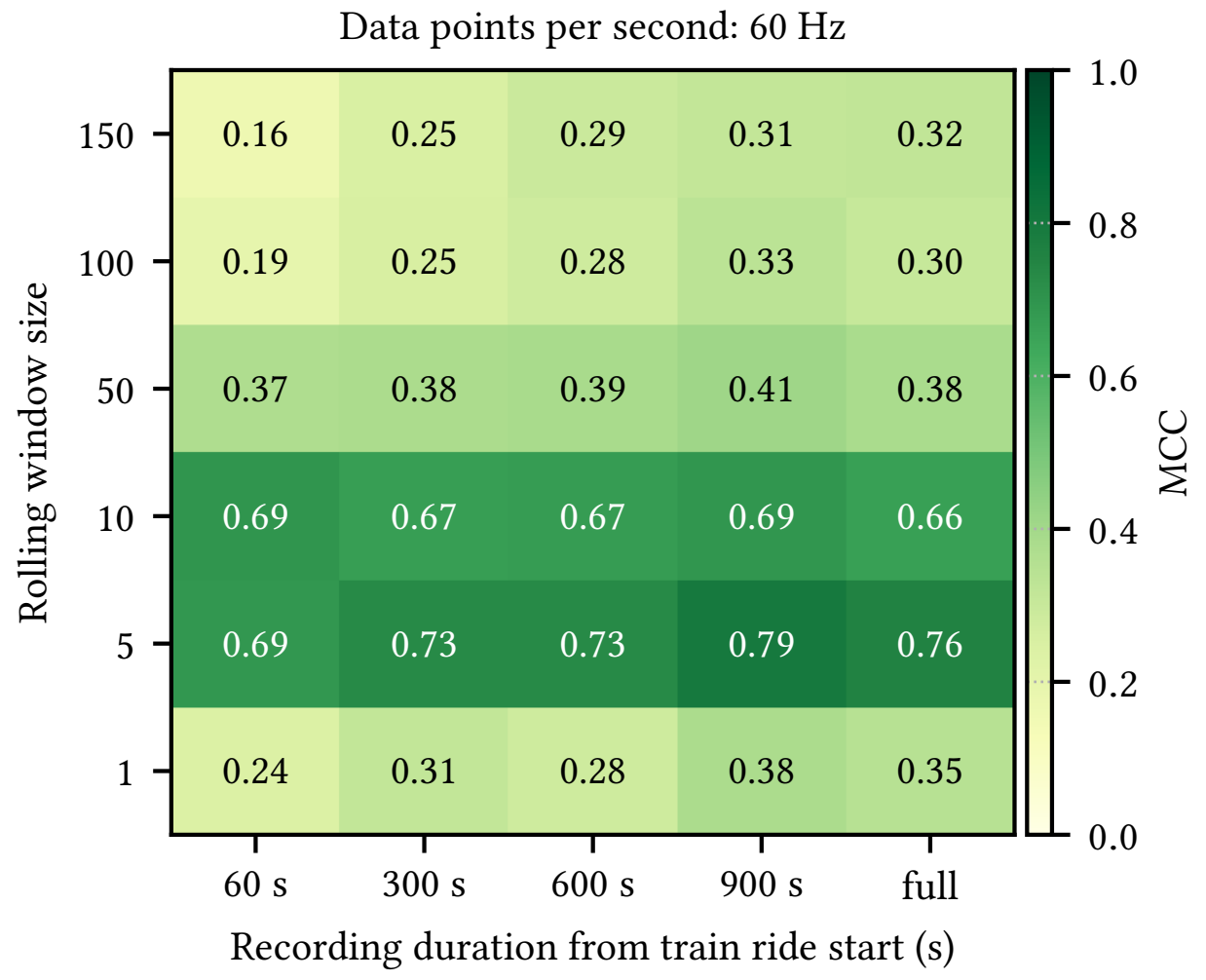}
    \caption{Learning-based colocation prediction performance as MCC score for different recording durations and rolling windows. The best MCC score is at a recording length of \MlResultBestFDuration\,s and a rolling window size of \MlResultBestFWindow.}
    \label{fig:coloc_ml_heatmap}
\end{figure}
We train the \ac{CNN} to predict the colocation of two chunks from two traces as described in~\cref{sec:approach_colocation}. The model is trained using a binary cross-entropy loss function, which measures the discrepancy between the predicted value and the true label, namely if the pair is colocated or not. The optimization is performed using the Adam optimizer with a linear scheduler gradually increasing the learning rate for the first few epochs.
The network is implemented in PyTorch.
Instead of drawing a single random pair from the training set as with the distance-based approach, we draw batches of pairs from the dataset uniformly at random. %
We trained our model on a NVIDIA A40 GPU for up to 1000 epochs on our training set, which took on average 30 minutes.

\paragraph{Hyperparameter search.}
In order to obtain an optimal configuration for the network architecture and subset of our dataset which yields the highest MCC for training, we used Optuna to search over \numprint{1000}~trials with a genetic sampling algorithm, i.e. the \ac{NSGAII}.
We provide the results of this ablation study regarding network architecture and training hyperparameters in~\cref{app:hparam_search}.
Training time varies slightly depending on the training set configuration, i.e. a higher sampling rate for longer recording durations leads to more combinations of pairs and, therefore, to increased training time.

\paragraph{Evaluation}
\Cref{fig:coloc_ml_heatmap} presents the \ac{MCC} scores across subsets of our evaluation dataset consisting of \MlResultBestMCCChunk s chunks, shown as multiple heatmaps over varying preprocessing parameters.
The \ac{CNN} is trained on one subset of our dataset, i.e. \SI{60}{\second} chunks recorded at \SI{60}{\hertz} with a rolling window size of 5 over the full recording length.
As smoothing acts as a low-pass filter, each window size changes the frequency content of the trace, and the model performs well only on the frequency range it sees during training.

\paragraph{Majority voting}
\label{sec:majority_vote_ml}
\begin{figure}[htbp]
    \import{figures}{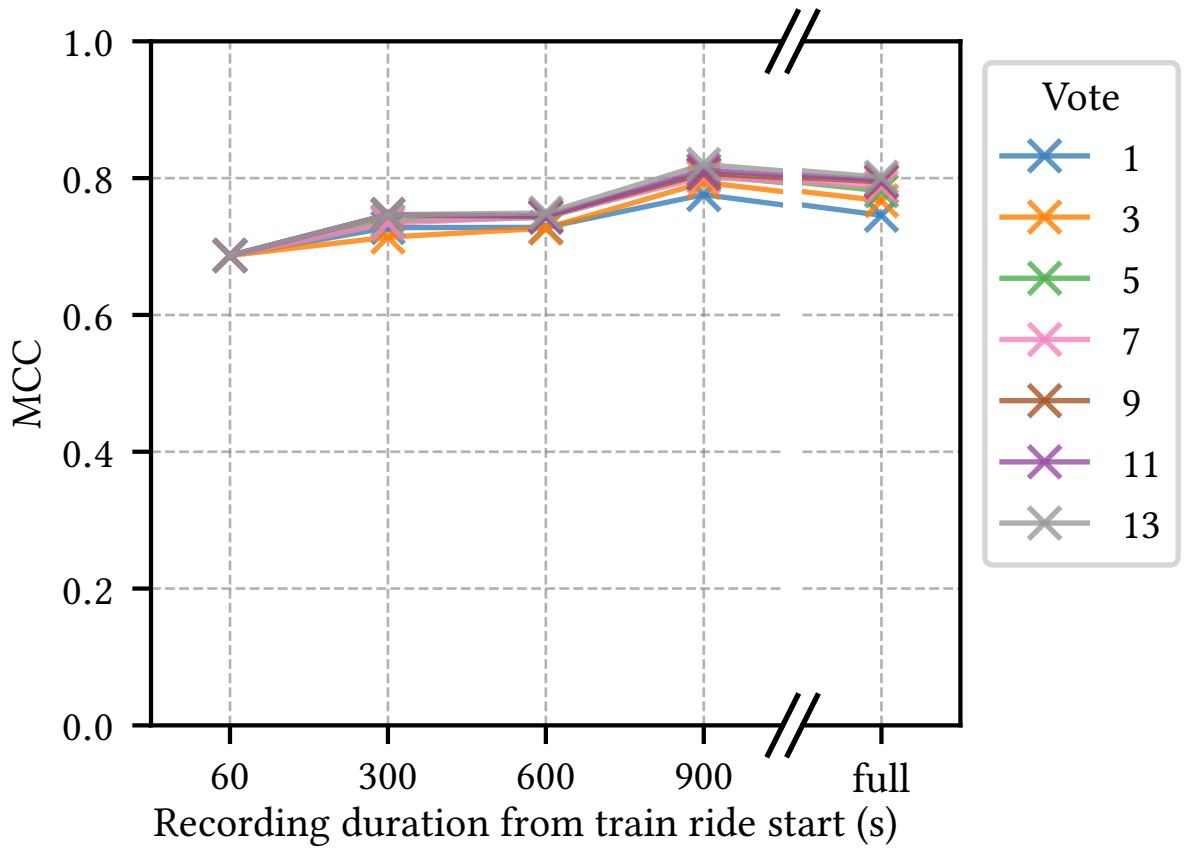}
    \caption{Learning-based colocation prediction evaluation with majority vote as MCC scores. Voting lengths $k$ are odd numbers up to $k=13$. From $k\geq5$, the MCC score does not improve significantly.}   
    \label{fig:majority_ml}
\end{figure}

We also analyze the impact of a majority vote (see~\cref{sec:tmd_majority}) on colocated, consecutive $k$ chunks from different devices.
In~\cref{fig:majority_ml} we present the \ac{MCC} scores for a chunk size of \SI{60}{\second}, a rolling window size of \MlResultBestFWindow, and data points recorded at a sampling rate of \SI{60}{\hertz}.
We can observe that the majority vote does not increase the \ac{MCC} score significantly after $k\geq5$. Nevertheless, given a larger recording duration, utilizing several consecutive chunks to predict colocation is beneficial.
Especially due to the very fast inference time, which we discuss in the following.

\subsection{Influence of the Train Type}
\label{sec:coloc_traintype}

Long-distance and regional services differ in speed, stopping frequency, and traction power.
The magnetic trace can therefore carry a different amount of information for each of them.
To test this, we repeat the evaluation of both variants on two subsets of our data, one with long-distance trips only and one with regional trips only.
\Cref{tab:coloc_traintype_combined} reports the MCC of both variants.

The two subsets differ in size, and shrink with the length of the post-departure window.
After \SI{300}{\second}, \ColocDatasetLongDistanceSegmentsFirstThreeHundred colocated segments remain for long-distance trips, but only \ColocDatasetRegionalSegmentsFirstThreeHundred for regional trips.
Regional trains stop more frequently, so few of their segments run longer than five minutes.
The regional subset falls to \ColocDatasetRegionalSegmentsFirstSixHundred segments after \SI{600}{\second} and to \ColocDatasetRegionalSegmentsFirstNineHundred after \SI{900}{\second}.
We therefore report no regional scores for these two windows, as there are not enough pairs on which to base them.

Long-distance trips match the full dataset closely, as they dominate our dataset and thus drive its overall result.
Regional trips reach higher scores in both variants, with \ColocDistanceRegionalDTWROneBestMCCFirstThreeHundred for the distance-based and \MlResultRegionalBestMCCOverall for the learning-based variant.
We see two possible reasons for this gap.
First, regional trains accelerate more often and more strongly, so each ride adds more characteristic events to the trace.
Second, the regional subset is much smaller, and since we compare each trace against all others, it offers fewer candidate pairs and thus fewer chances for false matches.
Given the large difference in subset size, we expect the second effect to dominate.

\begin{table}[htbp]
  \centering
  \caption{Colocation inference performance (MCC score) of the distance-based and machine learning approaches across different train types and initial time windows. Best ML performances per duration are highlighted in bold.}
  \label{tab:coloc_traintype_combined}
  \resizebox{\columnwidth}{!}{
  \begin{tabular}{ll ccc}
    \toprule
    \multirow{2}{*}{\textbf{Approach}} & \multirow{2}{*}{\textbf{First (s)}} & \multicolumn{3}{c}{\textbf{Train Type}} \\
    \cmidrule(lr){3-5}
    & & \textbf{All} & \textbf{Long distance} & \textbf{Regional} \\
    \midrule
    \textbf{Distance-based} & 60  & \ColocDistanceDTWROneBestMCCFirstSixty & \ColocDistanceLongDistanceDTWROneBestMCCFirstSixty & \ColocDistanceRegionalDTWROneBestMCCFirstSixty \\
                           & 300 & \ColocDistanceDTWROneBestMCCFirstThreeHundred & \ColocDistanceLongDistanceDTWROneBestMCCFirstThreeHundred & \ColocDistanceRegionalDTWROneBestMCCFirstThreeHundred \\
                           & 600 & \ColocDistanceDTWROneBestMCCFirstSixHundred & \ColocDistanceLongDistanceDTWROneBestMCCFirstSixHundred & - \\
                           & 900 & \ColocDistanceDTWROneBestMCCFirstNineHundred & \ColocDistanceLongDistanceDTWROneBestMCCFirstNineHundred & - \\
    \midrule
    \textbf{Machine Learning} & 60  & \MlResultBestMCCFirstSixty & \MlResultLongDistanceBestMCCFirstSixty & \textbf{\MlResultRegionalBestMCCFirstSixty} \\
                           & 300 & \MlResultBestMCCFirstThreeHundred & \MlResultLongDistanceBestMCCFirstThreeHundred & \textbf{\MlResultRegionalBestMCCFirstThreeHundred} \\
                           & 600 & \MlResultBestMCCFirstSixHundred & \MlResultLongDistanceBestMCCFirstSixHundred & \textbf{\MlResultRegionalBestMCCFirstSixHundred} \\
                           & 900 & \MlResultBestMCCFirstNineHundred & \textbf{\MlResultLongDistanceBestMCCFirstNineHundred} & \MlResultRegionalBestMCCFirstNineHundred \\
    \bottomrule
  \end{tabular}
  }
\end{table}

Overall, colocation inference works for both service categories and is therefore of little concern for the attacker.
The first five minutes of a ride are sufficient for both subsets to achieve a high success rate. 
Whether regional trains are in fact easier targets remains open, as the regional subset is too small to separate the effect of the train type from the effect of the subset size.

\subsection{Runtime Analysis}
\label{sec:runtime_analysis}

To assess whether each attack is computationally feasible, we measure the runtime of a single comparison between two recordings.
We evaluate all \numprint{114} colocation datasets, covering every combination of recording length and sampling rate, and measure \numprint{50} randomly selected recording pairs per dataset, repeated five times.
Only the comparison itself is measured, and loading and preprocessing are excluded.
To ensure a fair comparison, both approaches run on the same server using only the CPU, in our case an AMD EPYC 7713.

For the DTW-based approach a comparison is a single distance computation over the two recordings, whereas for the learning-based approach it is the classification of all chunk pairs of the two recordings followed by the majority vote, of which the vote accounts for at most \SI{11}{\percent} of the runtime.
The runtime of the learning-based approach is independent of the sampling rate, as the network resamples every chunk to a fixed input length.
Classifying a single chunk pair with the \ac{CNN} costs only \RuntimeMlPerChunkAmortised.
The cost of a single comparison is therefore dominated by the fixed cost of invoking the network once, not by the number of chunks to classify.
The classification of the whole trace grows with the recording length, from \RuntimeMlMin to \RuntimeMlMax.
In contrast, the DTW-based approach scales linearly with both factors, ranging from \RuntimeDtwMin to \RuntimeDtwMax.

\begin{figure}[htb]
    \import{figures}{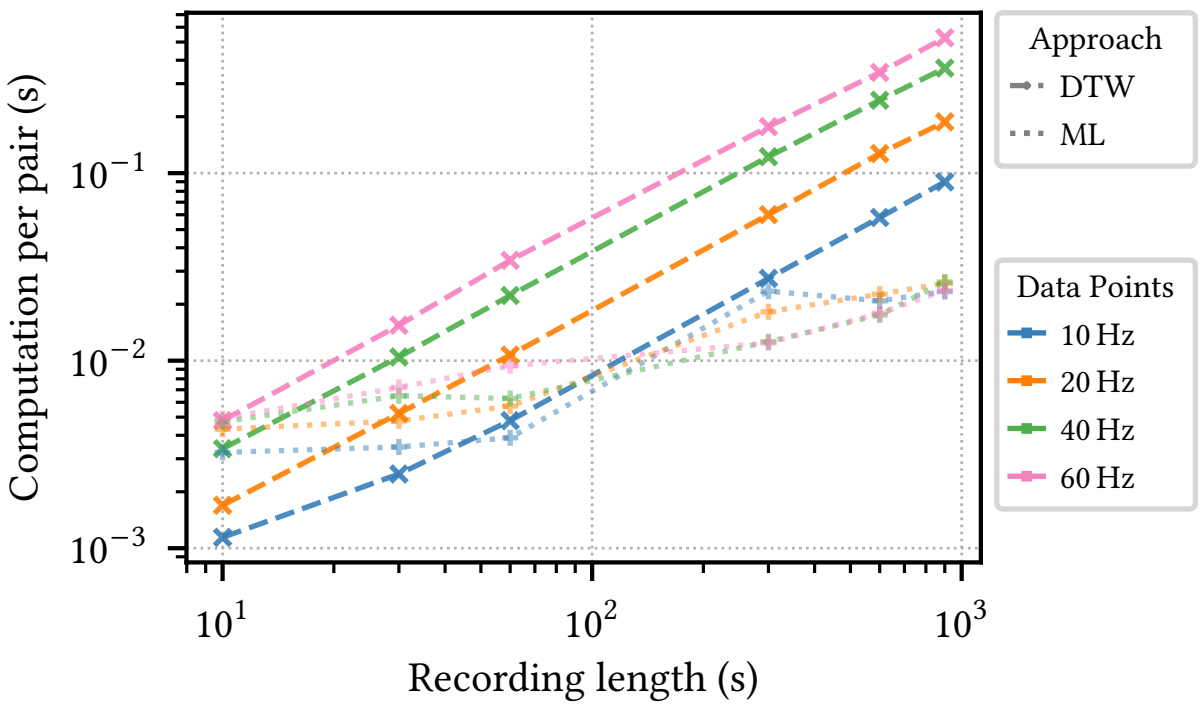}
    \caption{Runtime per comparison for the learning-based and distance-based colocation approaches.}
    \label{fig:runtime_analysis}
\end{figure}

\Cref{fig:runtime_analysis} illustrates the runtime of both approaches as a function of recording length. %
Averaged over the whole grid the learning-based approach is a factor of \RuntimeSpeedupMean faster than the DTW-based approach, rising to a factor of \RuntimeSpeedupMax for the longest recordings at the highest sampling rate.
This advantage can be increased further by using GPUs or other AI accelerators for inference.

\subsection{Summary}
\label{sec:coloc_summary}

\begin{table}[htbp]
\centering \small
\caption{
Colocation prediction performance of our attack and the related approach by Nguyen et al.}
\label{tab:coloc_comparison}
\begin{tabular}{@{}lrll@{}}
\toprule
\multicolumn{1}{c}{\textbf{Approach}}                                                                                                & \multicolumn{1}{c}{\textbf{First}} & \multicolumn{1}{c}{\textbf{F1}}             & \multicolumn{1}{c}{\textbf{MCC}}              \\ \midrule
\multirow{4}{*}{\begin{tabular}[c]{@{}l@{}}\citet{nguyenColocationEpidemicTracking2017}\\ reimplemented baseline\\ DDTW on magnitude\end{tabular}} & 60 s                                  & \ColocDistanceAbsDDTWBestFFirstSixty        & \ColocDistanceAbsDDTWBestMCCFirstSixty        \\
                                                                                                                                     & 300 s                                 & \ColocDistanceAbsDDTWBestFFirstThreeHundred & \ColocDistanceAbsDDTWBestMCCFirstThreeHundred \\
                                                                                                                                     & 600 s                                 & \ColocDistanceAbsDDTWBestFFirstSixHundred   & \ColocDistanceAbsDDTWBestMCCFirstSixHundred   \\
                                                                                                                                     & 900 s                                 & \ColocDistanceAbsDDTWBestFFirstNineHundred  & \ColocDistanceAbsDDTWBestMCCFirstNineHundred  \\ \midrule
\multirow{4}{*}{\begin{tabular}[c]{@{}l@{}}Our attack with\\ DTW distance\\ on full 16.7\,Hz trace\end{tabular}}                     & 60 s                                  & \ColocDistanceDTWROneBestFFirstSixty        & \ColocDistanceDTWROneBestMCCFirstSixty        \\
                                                                                                                                     & 300 s                                 & \ColocDistanceDTWROneBestFFirstThreeHundred & \ColocDistanceDTWROneBestMCCFirstThreeHundred \\
                                                                                                                                     & 600 s                                 & \ColocDistanceDTWROneBestFFirstSixHundred   & \ColocDistanceDTWROneBestMCCFirstSixHundred   \\
                                                                                                                                     & 900 s                                 & \ColocDistanceDTWROneBestFFirstNineHundred  & \ColocDistanceDTWROneBestMCCFirstNineHundred  \\ \midrule
\multirow{4}{*}{\begin{tabular}[c]{@{}l@{}}Our attack with\\ CNN model\\ on \MlResultBestMCCChunk\,s chunks of 16.7\,Hz trace\end{tabular}}               & 60 s                                  & \MlResultBestFChunkSixtyDurationSixty       & \MlResultBestMCCChunkSixtyDurationSixty       \\
                                                                                                                                     & 300 s                                 & \MlResultBestFChunkSixtyDurationThreeHundred & \MlResultBestMCCChunkSixtyDurationThreeHundred \\
                                                                                                                                     & 600 s                                 & \MlResultBestFChunkSixtyDurationSixHundred  & \MlResultBestMCCChunkSixtyDurationSixHundred  \\
                                                                                                                                     & 900 s                                 & \textbf{\MlResultBestFChunkSixtyDurationNineHundred} & \textbf{\MlResultBestMCCChunkSixtyDurationNineHundred} \\ \bottomrule
\end{tabular}
\end{table}

\Cref{tab:coloc_comparison} summarizes the colocation prediction performance of the baseline, the distance-based approach, and the learning-based approach on our dataset.
The baseline by~\citet{nguyenColocationEpidemicTracking2017} reaches an \ac{MCC} of 0.09 on our dataset, which is barely above chance.
This gap to the original study is consistent with the harder evaluation setting on traces sharing the same length introduced in \cref{sec:coloc_baseline}.
Our distance-based approach improves on the baseline by extracting the \SI{16.7}{\hertz} component before computing the \ac{DTW} distance, reaching a moderate \ac{MCC} of 0.60.
The learning-based approach performs best, with an \ac{MCC} of \MlResultBestMCCOverall.
Replacing the distance measurement with a \ac{CNN} lifts the score by a clear margin, on both long-distance and regional services, as shown in \cref{sec:coloc_traintype}.
The model is trained on a single subset of our data, yet it holds this margin across all recording durations.
It depends, however, on the smoothing used during training, so an attacker must preprocess new traces in the same way, which is easy to ensure in practice.

Both of our approaches reach their best score on the full length of the first \SI{900}{\second} of a ride.
But already after five minutes of travel, the learning-based approach can link two devices with only \MlResultBestMCCHz data points per second of the extracted trace.
This rate refers to the trace, not to the raw magnetometer, which still has to be sampled fast enough to resolve the \SI{16.7}{\hertz} component.

The two approaches also differ in computational cost.
A single pairwise \ac{DTW} computation in the best configuration of our distance-based approach takes about \SI{200}{\milli\second} on our system, while the learning-based approach needs below \SI{10}{\milli\second} for the same decision.
The runtime of \ac{DTW} additionally grows with both the rate and the length of the recording, while the runtime of the \ac{CNN} is independent.
Since the one-versus-all evaluation requires a quadratic number of such comparisons, this difference decides whether the attack scales to the $1{:}n$ setting of our threat model.
Overall, the learning-based approach is the only variant that combines high detection performance, robustness across train types, and the computational efficiency required for realistic large-scale deployment.

\section{Discussion}
We now place our findings in context and discuss what they mean for smartphone users.
The discussion is organized in four parts.
We first describe the privacy implications of our attacks, then state the assumptions under which they hold.
We then propose recommendations for mobile operating systems and close with a note on beneficial applications of the same signal.

\subsection{Implications}
Our results show that the smartphone magnetometer is a privacy-relevant sensor on electrified trains.
With our \ac{TMD} approach, an attacker can reliably tell whether a victim is on a train.
Once a train ride is detected, the same attacker can compare magnetic traces between devices and infer whether two victims share the same train.
Together, these two capabilities expose travel patterns and social proximity without any explicit location permission.
This places the magnetometer alongside the accelerometer and gyroscope as a sensor that warrants careful treatment in the mobile threat model.
The population most exposed by our specific instantiation are commuters in Central Europe, where the \SI{16.7}{\hertz} traction system is widespread and where train travel is part of daily routines.

\subsection{Assumptions}
Our attack rests on a small set of physical and technical assumptions.
First, the attack only applies to electrified trains and not to other modes of transport.
Second, it requires a railway network with a traction frequency that is extractable from a smartphone magnetometer.
Most current smartphones sample the magnetometer at 100\,Hz, and some at 125\,Hz.
Both support extraction of \SI{25}{\hertz} traction signals, and we see no physical barrier to transferability in such networks.
As sensor hardware improves, reliable measurement of 50\,Hz components may become possible, which would extend our approach to a large part of the world's rail network.
However, \SI{50}{\hertz} often coincides with the public power grid frequency, so the signal is no longer characteristic of railways alone.
In its current form, the geographic confinement of \SI{16.7}{\hertz} systems is itself an information leak, as the presence of this frequency in a trace reveals the continent on which the victim is located.

\subsection{Recommendations}
Two complementary mitigations are available at the operating system level.
The first is a stricter access control model for motion and environmental sensors.
At present, the magnetometer can typically be read by any application without explicit user permission, which enables continuous background recording.
As already suggested by others \cite{mohamedISTELANDisclosingSensitive2023,petraccaEnTrusTRegulatingSensor}, introducing a dedicated permission for the magnetometer and similar sensors would increase transparency and let users decide which applications may collect such data.
This would also make malicious applications easier to identify.

The second mitigation is a cap on the sampling rate exposed to applications.
Our attack relies on a stable \SI{16.7}{\hertz} component, which requires a temporal resolution of at least 35\,Hz in the recorded signal.
If the operating system limits the sampling rate available to applications below this value, frequency extraction is no longer possible.
Such a cap would directly remove the basis of both the \ac{TMD} and the colocation inference, but may also influence other features like step counting or gesture detection.

The two mitigations can be combined, by tying the available sampling rate to a permission tier.
A less restrictive variant is to require a special permission only when an application requests sampling rates that enable detailed spectral analysis.
This keeps benign applications working while denying malicious ones the resolution needed for frequency-based inference.

\subsection{Beneficial Applications}
The same signal that enables the attack can also support useful features.
Our frequency extraction provides a lightweight indicator for train rides that can be embedded into existing \ac{TMD} pipelines on smartphones, as current implementations have difficulties to distinguish between trains and cars (see \cref{sec:related_work}).
Most modern operating systems already offer activity recognition, and a robust train indicator would complement these features in a privacy-preserving and responsible deployment.
A second class of applications is colocation-based contact tracing.
During the COVID-19 pandemic, many contact tracing approaches were deployed and evaluated \cite{quachEffectivenessDigitalContact2025}.
The proposal of \citet{nguyenColocationEpidemicTracking2017} aligns with our setting, which makes colocation inference based on magnetic fields a candidate signal for contact tracing within trains.

\section{Related work}
\label{sec:related_work}

Having presented our two attacks, we now place them in the context of prior work along two axes: transport mode detection and colocation inference.
We focus on studies that, like ours, rely on permissionless smartphone sensors, since this is the property that makes the attacks of practical concern.

\paragraph{Transport mode detection}
Research on \ac{TMD} has grown in step with the spread of smartphones and can be organized along three axes: the sensors used, the set of inferred classes, and the inference method.
Since our approach uses permissionless sensors and considers train journeys as a transport mode, we restrict our discussion to related work that meets these requirements.
Broader surveys are provided by \citet{sadeghianReviewEvaluationMethods2021} and \citet{gongDerivingPersonalTrip2014}.
A range of further studies relies on protected sensor data such as GPS, Wi-Fi, sound or camera input and falls outside our scope, but we list them for completeness \cite{miluzzoSensingMeetsMobile2008, reddyUsingMobilePhones2010, nitscheSupportingLargescaleTravel2014, widhalmTransportModeDetection2012, fengTransportationModeRecognition2013, manzonivincenzoTransportationModeIdentification2011, ansarilariAutomatedTransportationMode2015, ashqarSmartphoneTransportationMode2019, richozTransportationModeRecognition2020, drosouliTransportationModeDetection2021, manivannanFrameworkIdentificationHuman2022, azizEnhancingTransparencyTransport2024}.

\newcommand{\code}[1]{\emph{#1}}
\citet{hemminkiAccelerometerbasedTransportationMode2013} propose the first approach that relies solely on the accelerometer for \ac{TMD}. They show the feasibility of distinguishing the modes \code{Still}, \code{Walking}, \code{Bus}, \code{Train}, \code{Metro} and \code{Tram}, and report an F1 score of \numprint{0.70} for the \code{Train} class on their own dataset. \citet{iskanderovBreakingLimitsTransportation2020} extend the set of classes to \code{Running}, \code{Stair Up}, \code{Stair Down}, \code{Bus}, \code{Motorcycle}, \code{Car}, \code{Bike}, and \code{High-Speed-Train}, using a \ac{LSTM} model that reaches an F1 score of \numprint{0.83} for the combined \code{Train} and \code{High-Speed-Train} class on the HTC dataset. As a follow-up, \citet{wangTransportationModeDetection2022, wangDFTransDualFrequency2022} propose two machine learning approaches with a smaller set of 8~classes, reaching F1 scores of \numprint{0.79} and \numprint{0.83} on HTC and of \numprint{0.74}  on SHL. Similarly, \citet{xuAutomaticClassificationTransportation2024} and \citet{taherinavidAutomaticTransportationMode2024} apply machine learning on HTC to infer 11~transport modes. \citet{ibanezMobilitAppDeepLearningBased2025} combine \acp{CNN} with an \ac{LSTM} on their own dataset to infer 12~classes, achieving an F1 score of \numprint{0.66} for train rides. On the NOR-TMD dataset, \citet{skrettingFeatureImportanceNovel2025} evaluate feature importance across multiple sensors and statistical descriptors with several machine learning methods up to an F1 score of \numprint{0.93}.

In contrast to these approaches, we address a binary decision rather than a multi-class problem and rely on the magnetometer alone.
This narrower setting is enabled by the \SI{16.7}{\hertz} trace of electric trains and yields the F1 scores reported in \cref{sec:tmd_results} with a much simpler model and shorter recordings.

\paragraph{Inferring colocation}
In contrast to \ac{TMD}, only a few studies on colocation rely on permissionless smartphone sensors, and most of them are motivated by contact tracing during epidemics. The existing work can be divided into approaches that use machine learning and approaches that do not. Without machine learning, \citet{nguyenColocationEpidemicTracking2017} propose an approach that uses the smartphone magnetometer to record the amplitude of the ambient magnetic field on London's public transport network. Pairs of recordings are compared with the \ac{DDTW} distance, and a threshold on this distance decides whether two devices were on the same vehicle. \citet{jeongSmartphoneMagnetometerBasedDiagnostic2019} address indoor colocation with the same sensor and also use traces of the magnetic field magnitude, but compare them with the Pearson correlation coefficient.
Among learning-based approaches, \citet{kosasihColocationDetectionSystem2020, kosasihUnsupervisedLearningBasedSpatial2021} propose a convolutional autoencoder model evaluated on a dataset of about 24~minutes~\citep{kosasihColocationDetectionSystem2020}, reporting F1 scores between \num{0.65} and \num{0.89}~\citep{kosasihUnsupervisedLearningBasedSpatial2021}.

All of these approaches operate on the raw amplitude or magnitude of the magnetic field, in either general transport or indoor settings.
Our work differs in two respects: we isolate the \SI{16.7}{\hertz} traction signal that is specific to electric rail transport, and we evaluate colocation on hundreds of real-world train rides rather than on minutes of recordings.

\medskip
Across both threads of research, prior work has not treated permissionless magnetometer access as a privacy attack surface in the rail setting.
Our paper closes this gap by showing that a single, narrow frequency component of the magnetometer signal is sufficient for both transport mode detection and colocation prediction under realistic conditions.

\section{Conclusion}
\label{sec:conclusion}

In this paper, we show that the magnetometer in smartphones leaks information about a user's train ride.
The traction current in the overhead line of an electric train produces a distinctive magnetic field at the grid frequency, which is measurable in the carriage and is captured by common smartphone sensors.
From this signal, we derive two attacks on passenger privacy that operate without any explicit permission on Android and iOS: transport mode detection and colocation inference.

We evaluate both attacks on a dataset of \TrainTrackJourneys~train rides collected under realistic, in-the-wild conditions.
For transport mode detection, a single spectral feature at \SI{16.7}{\hertz} is sufficient to reach an F1 score of \TmdResultBestFOverall and an MCC of \TmdResultBestMccOverall, on par with multi-sensor pipelines from prior work.
For colocation inference, a lightweight convolutional model on the extracted magnetic trace reaches an F1 score of \MlResultBestFOverall and an MCC of \MlResultBestMCCOverall, outperforming prior magnetometer-based approaches by a clear margin.
These results add to the growing body of evidence that unrestricted access to inertial sensors exposes smartphone users to privacy leaks.
We argue that these sensors should no longer be treated as harmless and mobile operating systems should place their access behind an explicit permission.

\begin{acks}
We would like to thank all contributors to the dataset, Elias Burggraef for reimplementing the NOR-TMD baseline and Stef Lenk for creating \cref{fig:train}.
This work was supported by the European Research Council (ERC) under the consolidator grant MALFOY (101043410) and the German Federal Ministry of Research, Technology and Space (BMFTR) under the grant ``AIgenCY'' (16KIS2012).
\end{acks}

\bibliographystyle{ACM-Reference-Format}
\bibliography{bibliography}

\section{Ethical Considerations}
\label{sec:ethics}
Research into attacks carries the potential for misuse, which we weigh against the need to understand emerging threats and to improve the security of widely deployed systems.
We follow the principles of the Menlo Report, in particular beneficence, respect for persons, and justice, and design our study to minimize harm while maximizing societal benefit.
Our research institute does not operate an Institutional Review Board (IRB), as is common in Europe, so we instead follow established best practices to avoid harm to stakeholders and to protect personal data throughout the study.

\paragraph{Stakeholders}
Our work affects several groups with varying degrees of exposure and risk.
Participants of our data collection are affected directly, as their recordings reflect travel behavior and movement patterns.
Smartphone users at large are affected indirectly, since we show that permissionless magnetometer data reveals transport mode and colocation.
The research community and industry stakeholders, such as smartphone manufacturers and operating system developers, benefit from our findings, as these insights help to design improved privacy protections and sensor access controls.
Adversaries, in contrast, could misuse the presented techniques to infer user behavior without consent.

\paragraph{Privacy of participants}
Participation is voluntary and based on informed consent.
All data is processed and analyzed in a controlled environment.
For the public release, we pseudonymize stations and segments and convert absolute timestamps into relative ones.
These measures reduce, but do not eliminate, the risk of deanonymization, and we explicitly inform participants about this residual risk during the consent process.
Together, they mitigate harms such as the privacy loss of participants.

\paragraph{Decision}
Our work reveals a security and privacy threat posed by magnetometer data from smartphones, and its publication may allow adversaries to replicate or extend our techniques.
At the same time, it enables countermeasures such as permission-based sensor access and reduced sampling rates, protecting a large population of smartphone users.
From a respect-for-persons perspective, participant rights are preserved through informed consent and privacy-preserving data handling.
While the possibility of misuse cannot be fully eliminated, we believe that disclosure of the threat is necessary to enable effective safeguards.
We therefore conclude that the expected benefits outweigh the risks and that both conducting and publishing this research is ethically justified.
An extended version of this section is available in our public artifact repository: \url{https://zenodo.org/records/22662413?preview_file=ETHICS.md}

\section{Open Science}
\label{sec:open_science}

We support transparency, reproducibility, and reuse of our work by providing all relevant artifacts whenever possible.
Our work is based on three main artifacts required to evaluate our core contributions.
These are the dataset, the transport mode detection pipeline, and the colocation approach.
For more details refer to the Artifact Appendix.

\paragraph{Dataset}
We provide the dataset, which contains raw sensor recordings collected during our study.
This includes magnetometer data as well as additional sensors recorded on the smartphones.
A detailed description of the dataset is given in the Artifact Appendix.

\paragraph{Transport mode detection}
We provide all code required to reproduce our transport mode detection pipeline.
This includes scripts for dataset generation, processing and evaluation.
We also provide our implementation of the method by \citet{skrettingFeatureImportanceNovel2025}.

\paragraph{Colocation}
We provide all code required to reproduce our colocation approach.
This includes dataset construction, processing of the \ac{DTW}, model training, and evaluation.
We also provide a trained model used in our experiments, the code for training and the optimized hyperparameters of the network.

\section{Generative AI Usage}
GitHub Copilot and Anthropic Claude were used to assist in transforming research code originally developed in Jupyter notebooks into a structured, release-ready code library.
All AI-assisted code was reviewed, tested, and validated by the authors through manual inspection.
To improve the understandability of the core library, GitHub Copilot was additionally used for documentation of the published code library.
All generated documentation was reviewed and corrected by the authors to ensure accuracy and consistency with the actual implementation.
OpenAI ChatGPT and Anthropic Claude were further used to check and improve the grammar, spelling and fluency of the submission.
All changes were reviewed, edited, and verified afterward by the authors.

\appendix
\section{Transport Mode Detection}
\label{app:tmd}

\subsection{NOR-TMD implementation}
\label{app:nor_tmd}
To implement the transport mode detection model originally proposed in the reference study, we re-created the preprocessing strategy described in the paper.
We applied this strategy to the dataset provided by the authors and chose the XGBoost algorithm for classification as the paper listed it to be the best-performing algorithm on the data by means of classification accuracy.

In our implementation, sensor recordings from mobile devices were first segmented into 10s time windows with an overlap of 5s.
Within each time window, we computed the described set of statistical aggregation features that summarize the behavior of each available sensor.
These features served as the input variables for the machine learning model.
For the model itself, we used the \ac{XGBoost} classifier from the XGBoost library.
The gradient boosting process sequentially constructs an ensemble of decision trees, where each new tree corrects the residual errors of the previous ones, leading to a robust classifier capable of capturing complex non-linear relationships between features and transport modes. 
Our reimplementation closely mirrors the methodology described in the original work.

\subsection{Detailed Results}
\label{app:tmd_results}

In addition to the highlighted results in \cref{sec:tmd}, the released artifacts provide the full list of evaluated parameter combinations with the results as CSV file.

\section{Colocation inference}

\subsection{Baseline Implementation}
\label{app:coloc_baseline}

The implementation of the approach by \citet{nguyenColocationEpidemicTracking2017} is straightforward, as described in \cref{sec:coloc_baseline}.
To establish this approach as a strong baseline, we evaluate multiple configurations of sliding window sizes and data point densities.
As with our approach, higher data point density leads to a slightly better performance, but in contrast, longer recordings degrade performance.
Furthermore, the consistently high standard deviation indicates that the DDTW-based approach lacks robustness across configurations.
In our released artifacts we provide the full list of evaluated parameter combinations with the results as CSV file.

\subsection{Hyperparameter Search and Network Architecture}
\label{app:hparam_search}

\begin{table}[htbp]
\centering
\caption{Best hyperparameters from the Optuna search after 200 epochs. A dash marks layers that are not present.}
\label{tab:optuna_hparams}
\small
\setlength{\tabcolsep}{5pt}
\begin{tabular}{lccccc}
\toprule
Trial & 940 & 991 & 792 & 952 & 693 \\
\midrule
MCC            & 0.728 & 0.728 & 0.716 & 0.715 & 0.714 \\
\addlinespace
Conv.\ layers  & 5 & 5 & 5 & 6 & 5 \\
Base channels  & 8 & 16 & 16 & 8 & 32 \\
Kernel size    & 7 & 7 & 7 & 7 & 5 \\
Stride         & 1 & 1 & 1 & 1 & 1 \\
Pool kernel    & 3 & 3 & 3 & 2 & 3 \\
Embedding dim  & 32 & 64 & 64 & 64 & 32 \\
\addlinespace
FC layers      & 3 & 3 & 3 & 1 & 3 \\
FC width 1     & 512 & 128 & 512 & 128 & 256 \\
FC width 2     & 64  & 128 & 128 & --  & 256 \\
FC width 3     & 64  & 256 & 128 & --  & 64  \\
\addlinespace
Dropout        & 0.269 & 0.269 & 0.315 & 0.374 & 0.269 \\
Learning rate  & 0.001 & 0.003 & 0.001 & 0.003 & 0.001 \\
Batch size     & 128 & 128 & 128 & 128 & 128 \\
\bottomrule
\end{tabular}
\end{table}

We tune both the network architecture and the training-related hyperparameters jointly using Optuna with the \ac{NSGAII} sampler over \numprint{1000} trials, as described in \cref{sec:coloc_ml}.
The full search space and the corresponding objective values for all trials are available in our published artifacts (see \cref{sec:open_science}). The five best trials and their results are reported in \cref{tab:optuna_hparams}.
\label{app:netarch}

The architecture of the 1D \ac{CNN} used for colocation inference is the result of the joint hyperparameter search, balancing detection performance against model complexity to keep inference cost low.
The complete layer-by-layer specification, including kernel sizes, number of filters, activation functions, and output dimensions, is provided as part of our published artifacts (see \cref{sec:open_science}).

\end{document}